\documentclass{article}

 \usepackage{graphicx}

\usepackage{amssymb}
 \usepackage{amsthm}
\usepackage{amsmath}
\usepackage{caption}
\usepackage{subcaption}
\usepackage{booktabs}
\usepackage{tabularx}
\usepackage{multirow}
\usepackage{placeins}
\usepackage{float}
\usepackage{setspace}
\usepackage{color}
\usepackage{nicefrac}
\graphicspath{{Figures/}}

\newcommand{\be} {\begin{equation}}
\newcommand{\ee} {\end{equation}}

\begin{document}

\title{SPARSE: A Subgrid Particle Averaged Reynolds Stress Equivalent Model: Testing with \textit{A Priori} Closure}

\author{
Sean L. Davis and Gustaaf B. Jacobs\\
Aerospace Engineering, San Diego State University, San Diego, CA\\
gjacobs@mail.sdsu.edu
\and
Oishik Sen and H.S. Udaykumar \\
Mechanical and Industrial Engineering, The University of Iowa, IO
}

%




\maketitle


\begin{abstract}
\it{
A Lagrangian particle cloud model is proposed that accounts for the effects of 
Reynolds-averaged particle and turbulent stresses and the averaged carrier-phase
velocity of the sub-particle-cloud scale on the averaged motion and velocity of the 
cloud. 
The SPARSE (Subgrid Particle Average Reynolds Stress Equivalent) model is based on a combination
of a truncated Taylor expansion of a drag correction
function and Reynolds averaging.  
It reduces the required number of computational parcels to trace a cloud of particles
in Eulerian-Lagrangian methods for the simulation of particle-laden flow .
Closure is performed in an \textit{a priori} manner 
using a reference simulation where all particles in the cloud are traced individually with a point particle model.
Comparison of a first-order model and SPARSE with the reference simulation
in one-dimension shows that both the stress and the averaging of the carrier-phase velocity on the cloud subscale
affect the averaged motion of the particle. A three-dimensional isotropic turbulence computation
shows that only one computational parcel is sufficient to  accurately trace a cloud of tens of thousand of
particles.
} 
\end{abstract}


\section{Introduction} 
\label{introduction}

The Eulerian-Lagrangian (EL) model, introduced by Crowe at al. \cite{CSS77, crowe_book}, is one of the major approaches used to for computing the interaction of a large number of particles with a turbulent flow.
In the EL model each particle is traced in its Lagrangian frame, i.e. the frame moving with the particle.
Because particles are treated as volumeless mathematical points in the EL ``point-particle'' approach, the tracing of many particles becomes much more computationally efficient.
With a point-particle assumption, the simulation of large number of particles in process-scale environments becomes feasible.

If the number of particles in a computation is prohibitively large, groups of physical particles are amalgamated into a single computational particle to further economize the computational cost.
This type of method is also known as Cloud-In-Cell (CIC) \cite{birdsall:JCP}.
In CIC methods, groups of particles are modeled as points and their motion is forced by the drag exerted on them by the fluid \cite{JD09, TD09, Davis1,
Davis2}.  
The CIC method as conventionally implemented does not account for sub-particle cloud dynamics
resulting from turbulent fluctuations or particle-particle interactions.

For the modeling of subgrid and/or sub-particle-cloud scales of turbulent carrier-phase flows, Large-Eddy Simulation 
is commonly employed.
In LES, filtered Navier-Stokes equations resolve large turbulent structures, while dissipation of energy from the subgrid scales (SGS) is modeled.
Commonly used SGS models include the classic Smagorinsky \cite{smagorinsky1} and Dynamic Smagorinsky \cite{dynamicsmagorinsky} models.

Similar to the modeling of SGS with eddy viscosity models in the carrier phase, SGS models are required for an accurate particle tracer \cite{Launder, Menter:1994, Sarkar90, Sutanu:1991, Sutanu:JFM}. 
To this end, models 
have been formulated with either a deterministic approach or stochastic modeling of subgrid scales.
In the deterministic approach, the instantaneous velocities are reconstructed for use in particle equations through defiltering \cite{Shotorban:2005, kuerten2005can, Shotorban:2005a, kuerten2006subgrid, shotorban2007improvement}.
Shotorban and co-workers \cite{Shotorban:2005, Shotorban:2005a, shotorban2007improvement} employed Approximate Deconvolution (AD) \cite{stolz2001approximate} for defiltering.
Deconvolution is a mathematical method to approximately reconstruct the instantaneous velocity through consecutively applying a filtering operation on the filtered velocities.
The consecutive application of the filtering operator is a result of a series expansion for the deconvolution.
Filtering, itself, is a convolution product of the instantaneous velocity and the filter kernel.
Kuerten and Verman \cite{kuerten2005can} employed a defiltering technique in which the filtering inversion is carried out in the Fourier space for the streamwise and spanwise directions while the inversion is approximated by a Taylor series for the cross-stream direction.
Although defiltering is efficient to implement, it can only be carried out for the represented modes.

A Langevin type stochastic differential equation can be used to compute the evolution of particles when the carrier phase is simulated by LES \cite{shotorban2004particle, Shotorban:2005, shotorban2006particle}.
A similar model was previously proposed for particles by Pozorski and Minier \cite{pozorski1998lagrangian, pozorski1999probability} and Minier and Peirano \cite{minier2001pdf} when the carrier phase is simulated by RANS.
Shotorban and Mashayek \cite{shotorban2006particle} extended  the application of the Langevin equation from a simple RANS framework \cite{pozorski1998lagrangian, pozorski1999probability, minier2001pdf} to a higher-resolution LES model by using the stochastic differential equation employed to solve LES equations through a Filtered Density Function approach \cite{gicquel2002velocity}.
A modified version of this model was proposed by Berrouk et al. \cite{berrouk2007stochastic} to account for crossing-trajectory effects.

While the modeling the effect of subgrid stresses on the dispersion of particles has been studied, the effect of sub-scale particle fluctuations under the computational particle assumptions of CIC has received much less attention.
Several articles \cite{chen1985, mostafa1988, tu1994} have studied the inclusion of small-scale particle-fluid energy transfer in their turbulence models.
However, these studies focus on the effect that the particles have on the fluid
rather than the influence of individual modeled particles on the averaged
computational particle dynamics.

In recent work, we have undertaken a multi-scaling modeling 
effort \cite{Sen15} in which macro-scale models such as the LES and CIC models are closed using results
from full resolution meso-scale simulations that solve for the turbulence and flow over particles
with moving boundaries. The general framework enables direct closure of averaging
terms in a wide parameter space that otherwise would take a great range of challenging and meticolous
experimentation to obtain emperical formulas. Within this framework, we have more freedom to develop
macro-models, for example, for particle cloud dynamics.

This paper presents a model that accounts for sub-scale interphase velocity perturbations on 
computational particle dispersion patterns in a CIC framework. The model is general and is a good
fit for a multi-scale framework.
The drag forcing term in the Lagrangian governing equation for the particle momentum employs a correction factor for high Reynolds and Mach numbers.
The particle drag correction factor is expanded using a Taylor expansion, which enables a simple Reynolds averaging of the governing equations to yield a second order perturbation term.
These higher order terms are deemed "Reynolds Stress Equivalent" terms, which capture the effects of meso-scale interphase velocity perturbations.
We therefore refer to the model as Subgrid Particle Averaged Reynolds Stress Equivalent, which leads to the appropriate acorynm SPARSE for this model
which requires a small number of computational particles to simulate
a larger number of real particles. 
The objective of this paper is to derive the model and demonstrate that SPARSE improves upon the tracing of the average location of a cloud of particles in CIC.
One-dimensional verification tests as well as a three-dimensional homogeneous turbulence flow case 
demonstrate the efficacy of the model.

The derivation of the SPARSE model is provided in the next section. 
In Section \ref{1d_testing}, we present \emph{a priori} testing of the SPARSE model on a one-dimensional analytic
 carrier-phase field followed by testing in a three-dimensional periodic box with decaying isotropic turbulence in Section \ref{3d_isotropic}.
Conclusions and recommendations are reserved for the final section.

\section{Derivation of the Model}
\label{derivation}
The Lagrangian equations that govern the particle motion under the point particle assumption in general form are,
\begin{eqnarray}
\frac{d{\bf {x_{p}}}}{dt} &= & {\bf {v_{p}}},\\
\label{eq:xp}
\frac{d{\bf {v_{p}}}}{dt} &= & \nicefrac{C_{D_s}}{\tau_p}({\bf u} - {\bf {v_{p}}}),
\label{eq:velp}
\end{eqnarray}
where ${\bf {v_{p}}}$ is the particle velocity vector,  ${\bf u}$ is the carrier-phase velocity vector.
and $\tau_p$ is the particle time constant \cite{crowe_book}.  The drag correction factor, $C_{D_s}$, is necessarily an empirical  function
that modifies the Stokes drag for a number of non-linear physical effects, such as Reynolds 
number \cite{crowe_book}, Mach number \cite{BKKPPF97}, or particle number density \cite{Tong}.  
Since  many of the physical corrections such as Reynolds number and Mach number correction
depend on the relative interphase velocity, ${\bf a}={\bf u} - {\bf {v_{p}}}$, we rewrite (\ref{eq:velp}) as
\begin{align}
\frac{d{\bf {x_{p}}}}{dt} &= {\bf {v_{p}}},\nonumber\\
\frac{d{\bf {v_{p}}}}{dt} &= f({\bf a})\cdot {\bf a}.
\label{gov_eqns_part_model}
\end{align}
to derive the SPARSE model. 
We have taken the correction function, $f({\bf a})=C_{D_s}/\tau_p$ that depends on the relative interphase velocity only. 
The derivation of the SPARSE model, however, is easily extended
for a correction function that is  dependent on more variables (see also Remarks below).


Cloud-In-Cell (CIC) models a group of particles with a single computational parcel. 
The properties of  the group of particles are ensemble averaged as 
\begin{align}
\overline{\eta} = \frac{1}{N}\sum_{i=1}^N \eta_i.
\end{align}
To derive a governing equation for the averaged single parcel motion
a Reynolds decomposition, $ \eta = \bar {\eta} + \eta^\prime$, is performed to split particle properties
into an averaged,   $\bar{\eta}$, and fluctuating, $ \eta^\prime$, component. Reynolds
averaging  (\ref{gov_eqns_part_model}) leads to
\begin{align}
\frac{d \overline{(\bar{\bf x}_{p} + {\bf x}_{p}^\prime)}}{dt} &= \overline{\bar{\bf v}_{p} + {\bf v}_{p}^\prime},\nonumber\\
\frac{d \overline{(\bar{\bf v}_{p} + {\bf v}_{p}^\prime)}}{dt} &= \overline{f(\bar{\bf a} + {\bf a}^\prime})\cdot (\bar{\bf a} + {\bf a^\prime}).
\end{align}
Using 
\begin{align}
\overline{\frac{d \eta}{dt}} = \frac{d \bar{\eta}}{dt},
\end{align}
 for linear  derivative operators,
the averaged computational particle velocity and position can be written as
\begin{align}
\frac{d{ \bar{\bf x}_{p}}}{dt} &= {\bar{\bf v}_{p}},\nonumber\\
\frac{d{ \bar{\bf v}_{p}}}{dt} &= \overline{f(\bar{\bf a} + {\bf a}^\prime)\cdot (\bar{\bf a} + \bar{\bf a}^\prime)}.
\end{align}
Here, $\bar{\bf a} = \bar{\bf u} -   \bar{\bf {v}}_{p}$ where $\bar{\bf u}$
 is the average of the carrier-phase velocity at all particle locations and 
${\bf a}^\prime = {\bf u}^\prime - {\bf  v}_{p}^\prime$ where ${\bf u}^\prime$ are the carrier-phase velocity
fluctuations at the particle positions and ${\bf v}_p^\prime$ are the fluctuations in
the particle phase.

In the tradtional CIC approach \cite{birdsall:JCP} the average of the correction factor term is simply set as the correction factor for the average interphase velocity,
\begin{equation}
\overline{f(\bar{\bf a} + \bf{a}^\prime}) = f(\bar{\bf a}),
\label{eq:firstorder}
\end{equation}
leading to 
\begin{align}
\frac{d \bar{\bf x}_{p}}{dt} &=   \bar{\bf v}_{p},\nonumber\\
\frac{d \bar{\bf v}_{p}}{dt} &= f(\bar{\bf a})\cdot (\bar{\bf a}),
\label{first_order_model}
\end{align}
where the averaged carrier-phase velocity is taken as the carrier-phase velocity
 at the average particle position, $\bar{\bf u} = {\bf u}(\bar{\bf x}_{p})$. 

Instead of the assumption in (\ref{eq:firstorder}), we propose to Taylor
expand, $f(\bar{\bf a} + {\bf a} ^\prime)$ around  ${\bar{\bf a}}$. With the three-dimensional vector components,
$\bar{\bf a}=(a_1,a_2,a_3) = (\bar{a}_1+a_1^\prime,\bar{a}_2+a_2^\prime,\bar{a}_3+a_3^\prime)$, this leads to
\begin{align}
f({\bf a}) = &f( \bar{\bf  a}) + \frac{df(\bar{\bf a})}{da_1}((\bar{a}_1+a_1^\prime)-\bar{a}_1)+ \frac{df(\bar{\bf a})}{da_2}((\bar{a}_2+a_2^\prime) 
-\bar{a}_2) ...\nonumber\\
&+ \frac{df(\bar{\bf a})}{da_3}((\bar{a}_3+a_3^\prime) -\bar{a}_3) + O({a^\prime}^2),\nonumber\\
= &f(\bar{\bf a}) + \frac{df(\bar{\bf a})}{da_1}a_1^\prime + \frac{df(\bar{\bf a})}{da_2}a_2^\prime+ \frac{df(\bar{\bf a})}{da_3}a_3^\prime+ O({a^\prime}^2).
\label{taylor_expansion}
\end{align}
The Taylor series is truncated at second order terms $O({a^\prime}^2)$ for interphase velocity fluctuations that
are small as compared to the averaged interphase velocity.

Substituting (\ref{taylor_expansion}) in the particle momentum equation (\ref{gov_eqns_part_model}) yields,
\begin{align}
\frac{d{\bf v}_{p}}{dt} = (f(\bar{\bf a}) + \frac{d f({\bf \overline a})}{d {\bf a}}{\bf a}^\prime)(\bar{\bf a}+{\bf a}^\prime).
\end{align}
After expanding and averaging, we obtain
\begin{align}
\frac{d\bar{\bf v}_{p}}{dt} = &f(\bar{\bf a})\bar{\bf a}
+ \frac{df(\bar{\bf a})}{d {\bf a}}\overline{{\bf a}^\prime {\bf a}^\prime} 
\label{eq:SPARSE}
\end{align}
In (\ref{eq:SPARSE}) the terms, $\overline{{\bf a}^\prime {\bf a}^\prime}$, are recognized as stresses that typically arise in Reynolds
averaging and that require closure. 
In the present study these terms will be closed using an \emph{a priori} approach where the 
results of Eulerian-Lagrangian simulations are averaged to find the stresses  in the SPARSE model.

\vspace{0.7cm}

{\large \textit{Remarks:}}

\renewcommand{\labelenumii}{\Roman{enumii}}
\begin{enumerate}
\item The average carrier-phase velocity in the cloud is not equal to the carrier-phase
      velocity at the averaged particle location, $\overline{{\bf  u}({\bf x}_p)} \neq
      {\bf u}(\bar{\bf x}_p)$. A model is required to close the averaged
      carrier-phase velocity of the particles. 
      In this paper we use an \textit{a priori} closure.
\item Equation (\ref{first_order_model}) results if in the SPARSE derivation  
      the first-order term in the Taylor expansion  in (\ref{taylor_expansion}) is truncated.
      The CIC is hence a first-order model, whereas  SPARSE is a second-order model. 
      Depending on the truncation of the Taylor series any order model may be derived.  
      In the remainder of this paper, we refer to the combination of the 
      CIC model and the carrier-phase velocity at averaged particle location taken 
      (erroneously) to be equal to  ${\bf u}(\bar{\bf x}_p)$ as a First-Order Model. 
      The combination of a second-order model and $\overline{u(\bf x)}$ as the carrier-phase velocity, we shall
      refer to as the SPARSE model.
\item Taylor expansion of the correction function leads to terms in the particle momentum equation that have integer powers only. 
      Reynolds averaging on terms with integer powers is straightforward, whereas Reynolds averaging on terms with real and fractional powers (typical for the empirical functions such as $C_{Ds}$ ) is very  challenging.
\item It is assumed that the interphase velocity perturbations are small. This is not always the case. For example,
      when the particle response time, $\tau_p$, is large (heavy particles) and the initial particle velocities are large and have
      a large variation, then the fluctuations of the particles are large compared to the computational particle velocity. 
      In this case the validity of CIC and/or SPARSE will only hold for short times.
\item If $f({\bf a})$ depends on variables other then the interphase velocity, for example, particle diameter or number density,
      then a Taylor expansion has to be performed around those variables also. Averaging leads to additional second-order correlations
      that require closure.
\item For a \textit{posteriori} tests, the interphase stress terms require modeling and extensive knowledge of 
      $f({\bf a})$. In \cite{Sen15} we present a multi-scale framework in which we use full-scale simulations 
      to obtain detailed information for $f({\bf a})$.  
\end{enumerate}

\section{One-dimensional Verification Tests}
\label{1d_testing}

The SPARSE model improves upon the First-Order model in two ways: Firstly, SPARSE
accounts for the subscale particle fluctuations through stress terms. Secondly, 
SPARSE averages the carrier-phase velocity at the particle locations correctly.
In this section, we verify the impact of these two model improvements with 
one-dimensional test cases.


In the tests,  the carrier field is analytically 
specified to prevent errors in the carrier-phase solver from polluting  the model error 
investigation. 
Ten thousand particles initialized in a cloud are individually traced  according to
(\ref{gov_eqns_part_model}). We refer to this solution as the ``exact'' solution. 
SPARSE and the First-Order model trace the averaged location and velocity of this cloud
with a single computational parcel. 
The stress terms and carrier-phase velocity at the particle location are \textit{a priori} obtained
from the ``exact'' solution.

\subsection{Effect of the Reynolds Stress Equivalent Terms}
\label{uniform_velocity}

To isolate the effect of the stress closure from the effect of the carrier-phase velocity closure, 
we first consider a constant carrier-phase velocity,
\begin{align}
u(x,t) = 10,
\end{align}
in which case $\overline{u(x_p)} = u(\bar{x}_p)$, i.e. the carrier-phase
velocity modeled by the First-Order model and the SPARSE model are the same in this case. The
difference between the two models can hence only be caused by the stress terms in SPARSE.

For the exact reference case ten thousand  particles are initialized at the following locations and velocities,
\begin{align}
x_p &= \sigma_x\cdot d_{cloud},\nonumber\\
v_p &= 5.0 + \sigma_v\cdot\gamma,
\label{eq:ic_1d}
\end{align}
where the cloud diameter, $d_{cloud}$=1, and $\sigma_x, \sigma_y$ are uniformly distributed random numbers
between -1 and 1.  The maximum amplitude of the initial velocity perturbations is $\gamma$=10.
The modeled parcel's location and velocity is initialized with the mean of the exact locations and velocities, respectively.
Time integration is performed using a first order explicit time stepping routine with a time step of $\Delta t = 10^{-5}$, which 
ensures that numerical time integration errors are smaller than the model errors.

\begin{figure}[h]
	\centering
	\begin{subfigure}[b]{0.48\textwidth}
		\includegraphics[width=\textwidth]{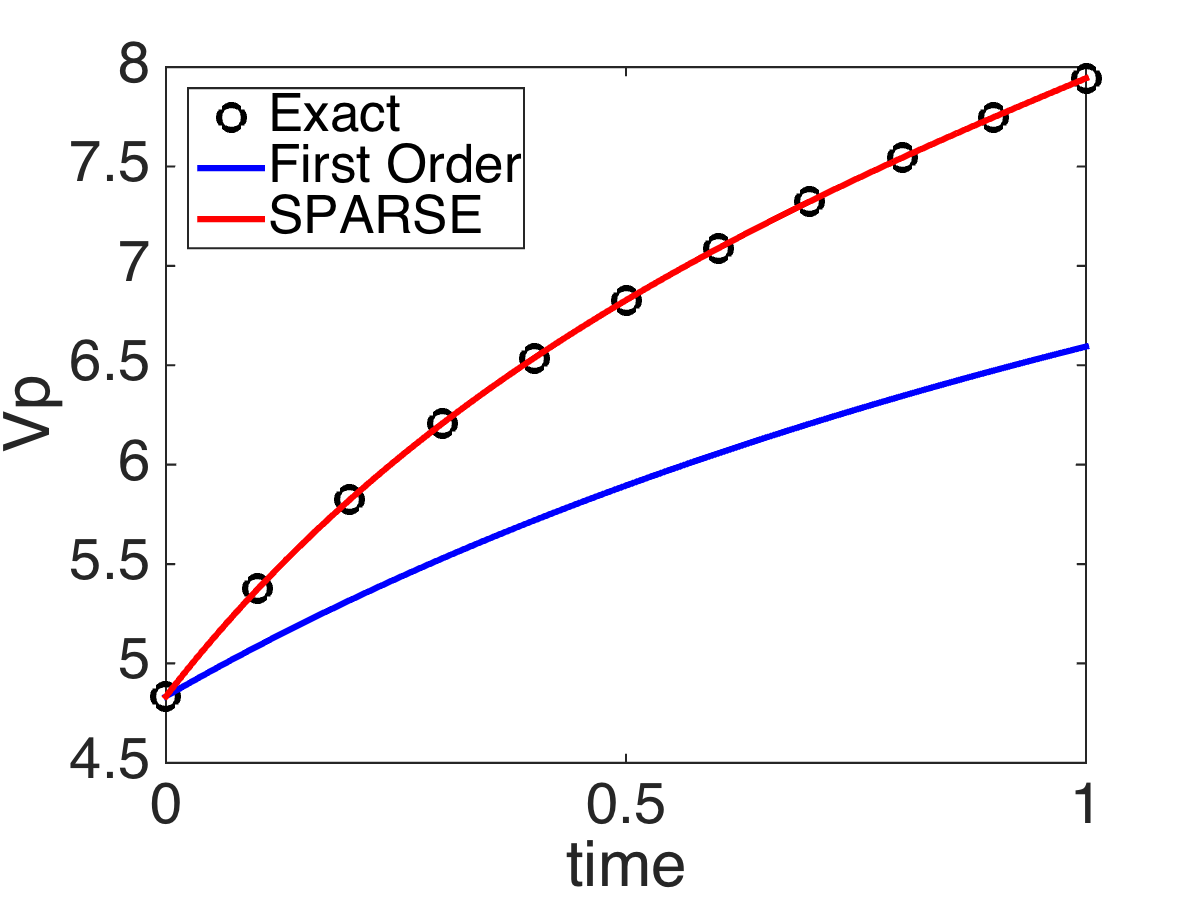}
		\caption{}
		\label{case_1_magnitudes_vp}
	\end{subfigure}
	\begin{subfigure}[b]{0.48\textwidth}
		\includegraphics[width=\textwidth]{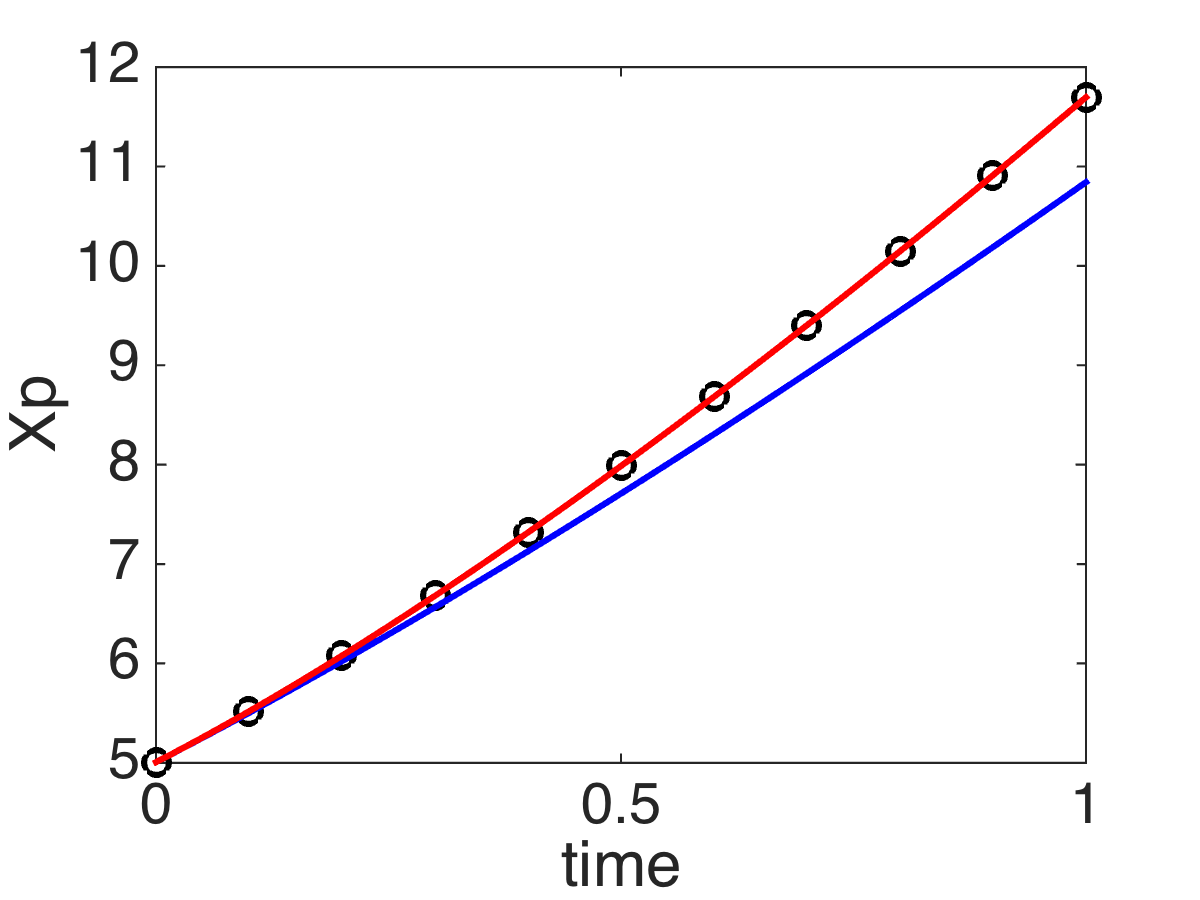}
		\caption{}
		\label{case_1_magnitudes_xp}
	\end{subfigure}
	\caption{Particle (a) velocities and (b) locations using the mean of the exact particle locations as well as the First-Order and SPARSE models with a uniform background fluid velocity and a linear correction factor.}
	\label{case_1_magnitudes} 
\end{figure}

\begin{figure}[h]
	\centering
	\begin{subfigure}[b]{0.48\textwidth}
		\includegraphics[width=\textwidth]{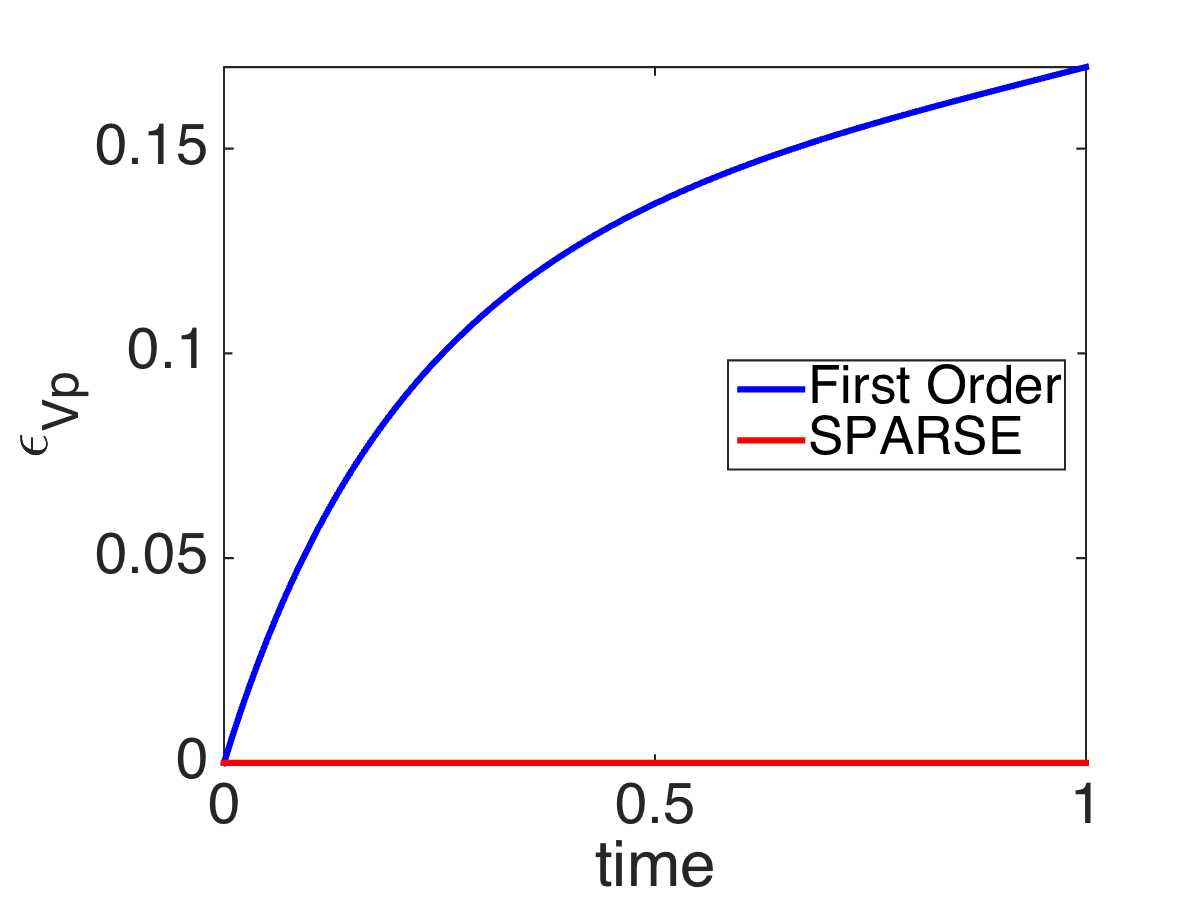}
		\caption{}
		\label{case_1_errors_vp}
	\end{subfigure}
	\begin{subfigure}[b]{0.48\textwidth}
		\includegraphics[width=\textwidth]{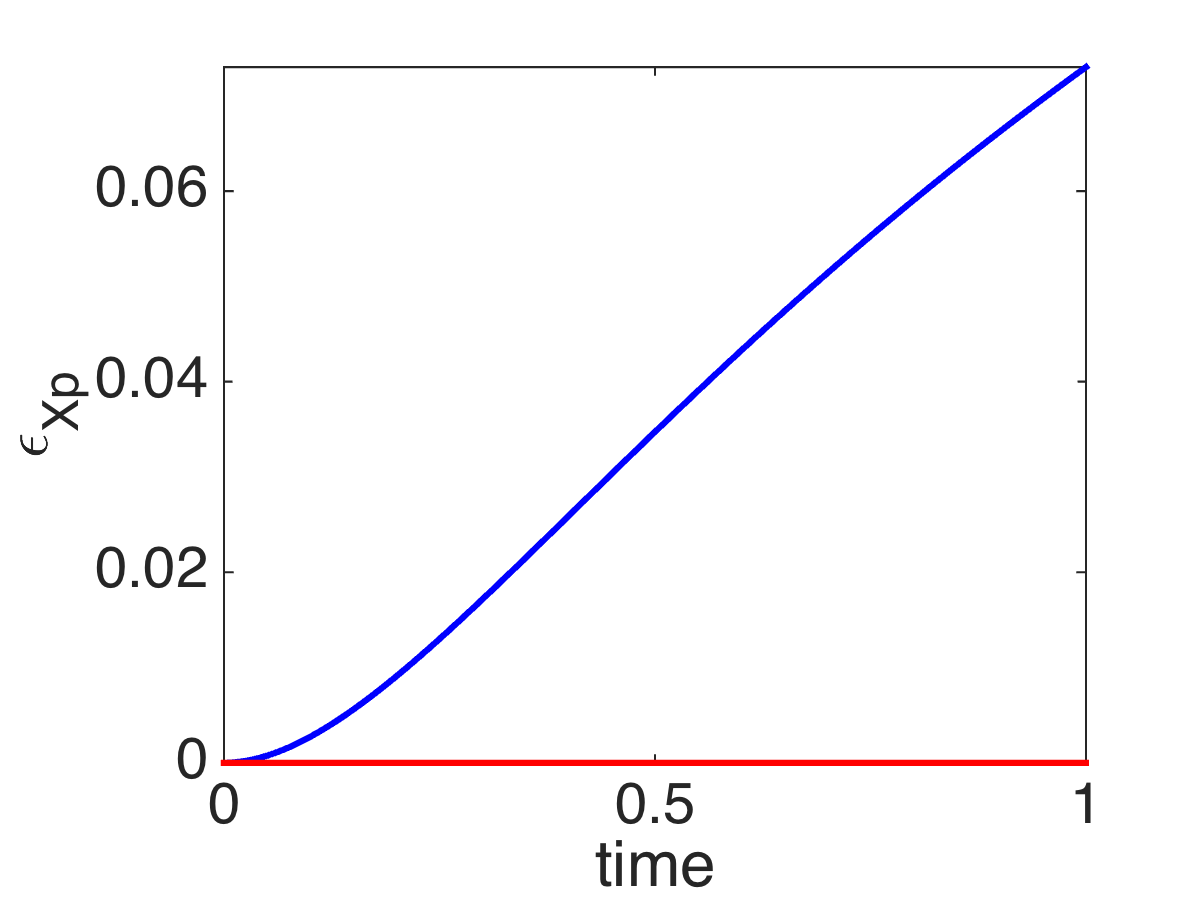}
		\caption{}
		\label{case_1_errors_xp}
	\end{subfigure}
	\caption{Modeling error of the mean particle (a) velocity and (b) location using the First-Order and SPARSE models with a uniform background fluid velocity and a linear correction factor.}
	\label{case_1_errors} 
\end{figure}

The particle correction factor, $f(a)$, is taken to  be a
linear function of $a$,
\begin{align}
f(a) = \frac{a}{10},
\label{eq:linearf}
\end{align}
similiar to low Reynolds number correction factors \cite{schiller_naumann}. 
Since derivatives of order higher than one are zero, 
the second-order expansion in (\ref{taylor_expansion})  and hence the SPARSE model is exact.  
With exact \textit{a priori} closure
the SPARSE model is the same as the exact model as confirmed in Figure \ref{case_1_magnitudes}.
Figures \ref{case_1_errors_vp} and \ref{case_1_errors_xp} show 
the errors in  the location and velocity in time determined with
\begin{align}
\epsilon_{x_p} = \frac{|\overline{x_{p,exact}} - x_{p,model}|}{\overline{x_{p,exact}}},\nonumber\\
\epsilon_{v_p} = \frac{|\overline{v_{p,exact}} - v_{p,model}|}{\overline{v_{p,exact}}}.
\label{SPARSE_errors}
\end{align}
The figures highlight the importance of inclusion of the second order stress term in SPARSE.
The truncation of the expansion at $O(a')$ in the First-Order model leads to errors in the averaged particle velocity 
on the order of tens of percentage (a maximum of 17\%) at time $t=1$ whereas the SPARSE model is exact.
The First-Order model does not account for the subscale kinetic energy in the particle phase, causing the modeled particle to lag behind the average particle position of the exact cloud.

For the initial conditions in (\ref{eq:ic_1d}), 
the interphase velocity difference can be negative and hence the correction factor in (\ref{eq:linearf})
can also have negative values. 
For a negative correction factor particles are nonphysically propelled.

Using the following correction factor of
\begin{align}
f(a) = \frac{|a|}{10},
\label{eq:fa_1dabs}
\end{align}
based on the absolute velocity, $|a|$, is more physical since $f(a)$ is always positive. 
Note that $f(a)$ in (\ref{eq:fa_1dabs}) is non-linear
because of discontinuous derivatives at $a=0$. 

\begin{figure}[h]
	\centering
	\begin{subfigure}[b]{0.48\textwidth}
		\includegraphics[width=\textwidth]{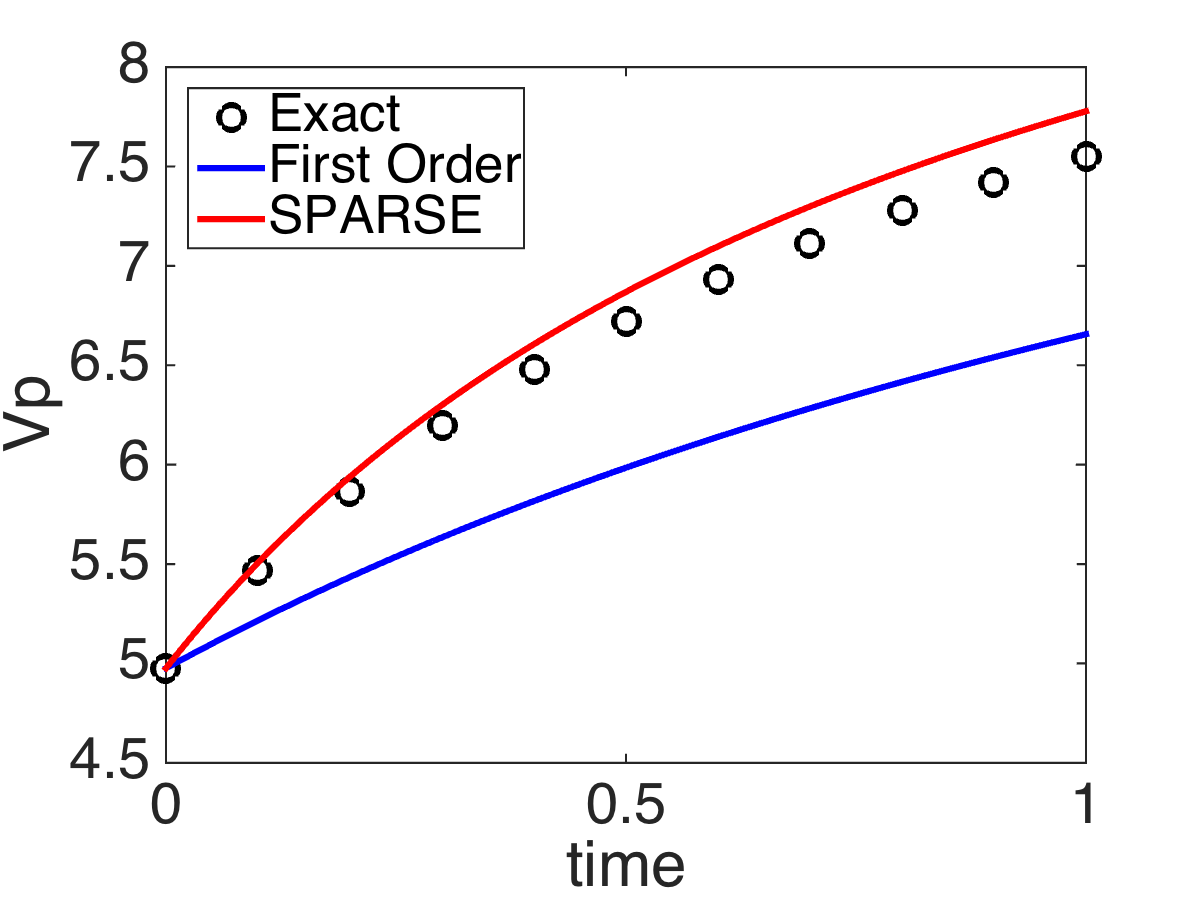}
		\caption{}
		\label{case_2_magnitudes_vp}
	\end{subfigure}
	\begin{subfigure}[b]{0.48\textwidth}
		\includegraphics[width=\textwidth]{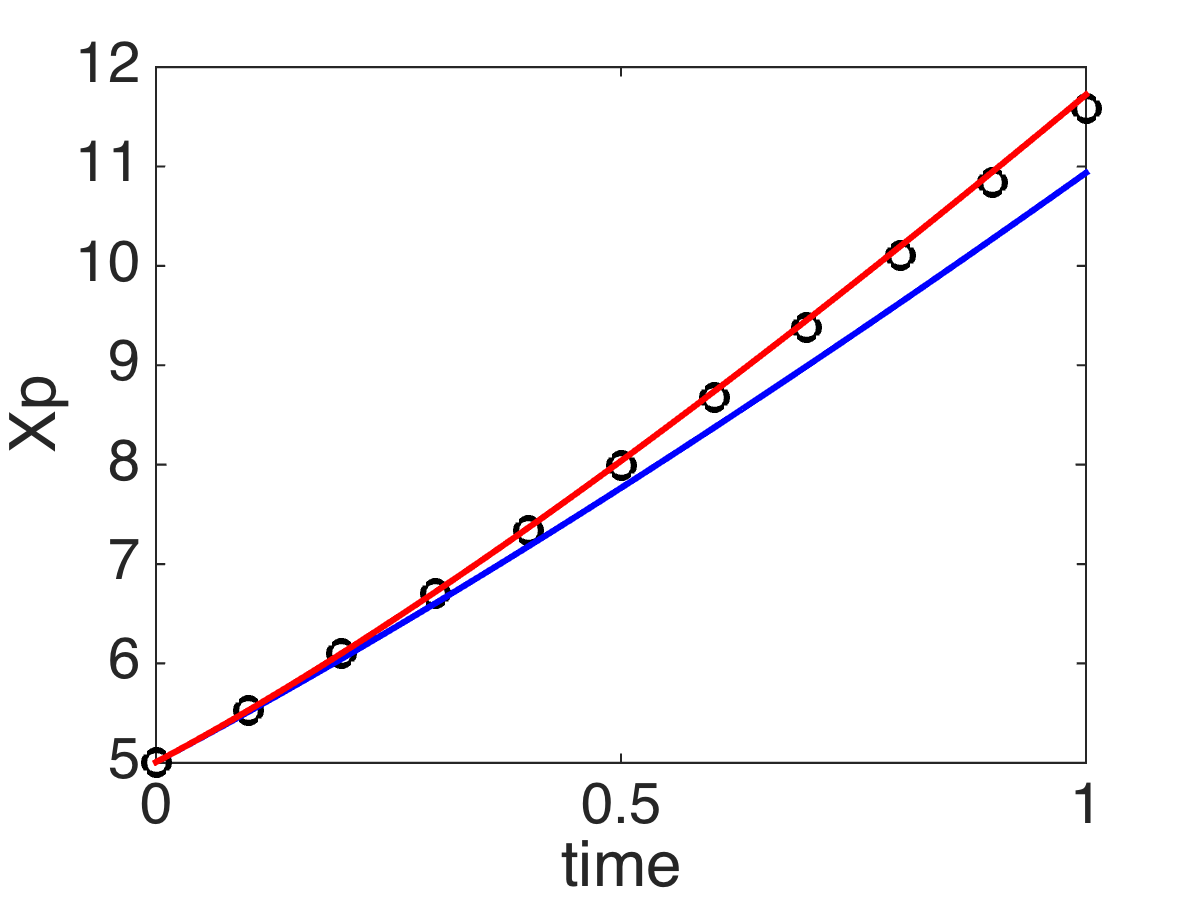}
		\caption{}
		\label{case_2_magnitudes_xp}
	\end{subfigure}
	
	\caption{Particle (a) velocities and (b) locations using the mean of the exact particle locations as well as the First-Order and SPARSE models with a uniform background fluid velocity and an absolute value correction factor.}
	\label{case_2_magnitudes} 
\end{figure}

\begin{figure}[h]
	\centering
	\begin{subfigure}[b]{0.48\textwidth}
		\includegraphics[width=\textwidth]{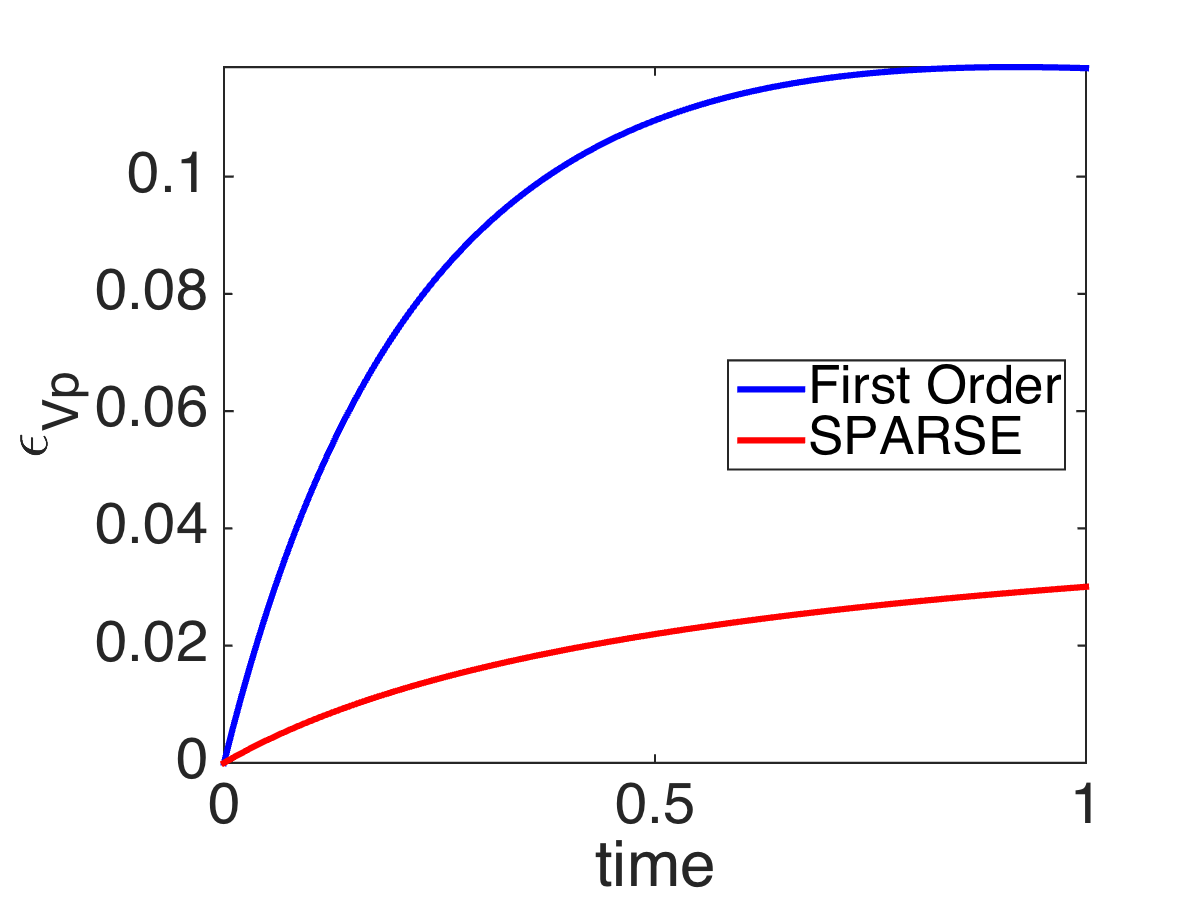}
		\caption{}
		\label{case_2_errors_vp}
	\end{subfigure}
	\begin{subfigure}[b]{0.48\textwidth}
		\includegraphics[width=\textwidth]{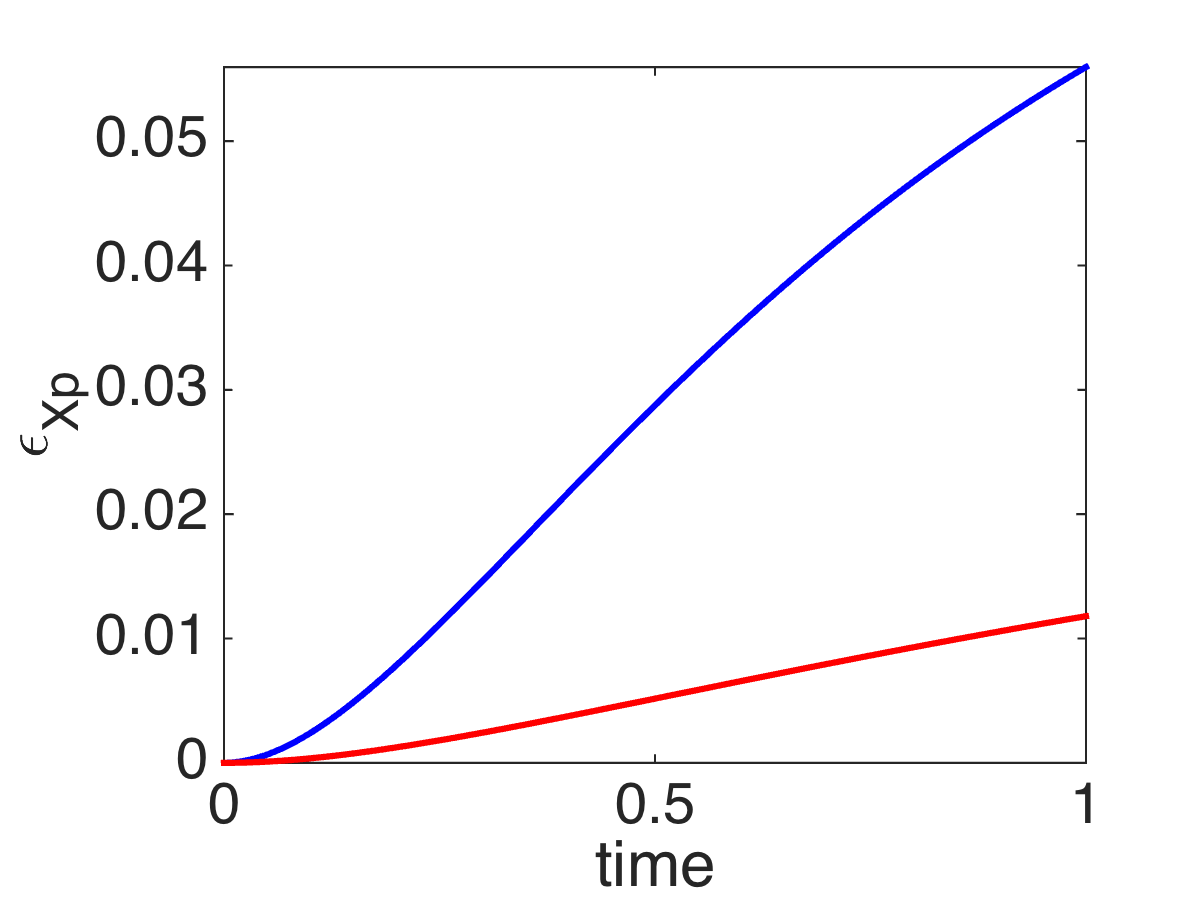}
		\caption{}
		\label{case_2_errors_xp}
	\end{subfigure}
	
	\caption{Modeling error of the mean particle (a) velocity and (b) location using the First-Order and SPARSE models with a uniform background fluid velocity and an absolute value correction factor.}
	\label{case_2_errors} 
\end{figure}

The non-linearity leads to errors in the modeling of the average particle velocity with the SPARSE model (Figures \ref{case_2_magnitudes_xp} and \ref{case_2_errors}).  
%
Because the SPARSE model accounts for the perturbations in the particle phase, 
it  is still significantly  more accurate than the First-Order model.

\subsection{Effect of Modeling the Average Cloud Velocity}
\label{uniform_correction_factor}

To test the erroneous assumption that $\bar{u}({x}_{p})$ = $\bar{u}(\bar{x}_{p})$
in the First-Order model,  a constant correction factor comparable to the Stokes' drag \cite{stokes} is used,
\begin{align}
f(a) = \frac{24}{St}.
\end{align}
The combination of the constant correction factor, for which the Taylor expansions terms are zero, and a spatially varying carrier-phase velocity,
\begin{align}
u(x) = \cos(\pi x) + x,
\end{align}
enables an investigation into the errors caused by the incorrect sampling of the carrier-phase velocity at a single point in the 
First-Order model.

\begin{figure}[ht]
	\centering
	\begin{subfigure}[b]{0.48\textwidth}
		\includegraphics[width=\textwidth]{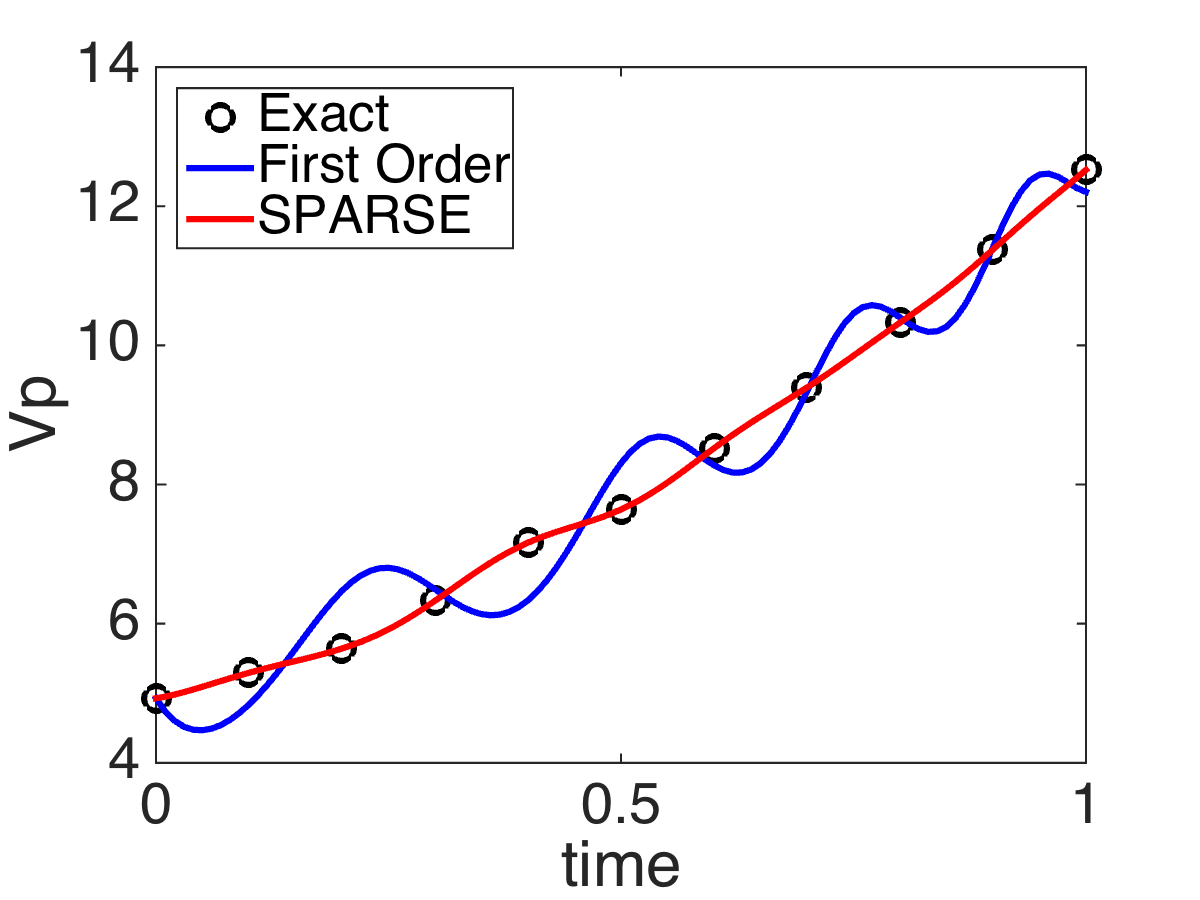}
		\caption{}
		\label{case_3_magnitudes_vp}
	\end{subfigure}
	\begin{subfigure}[b]{0.48\textwidth}
		\includegraphics[width=\textwidth]{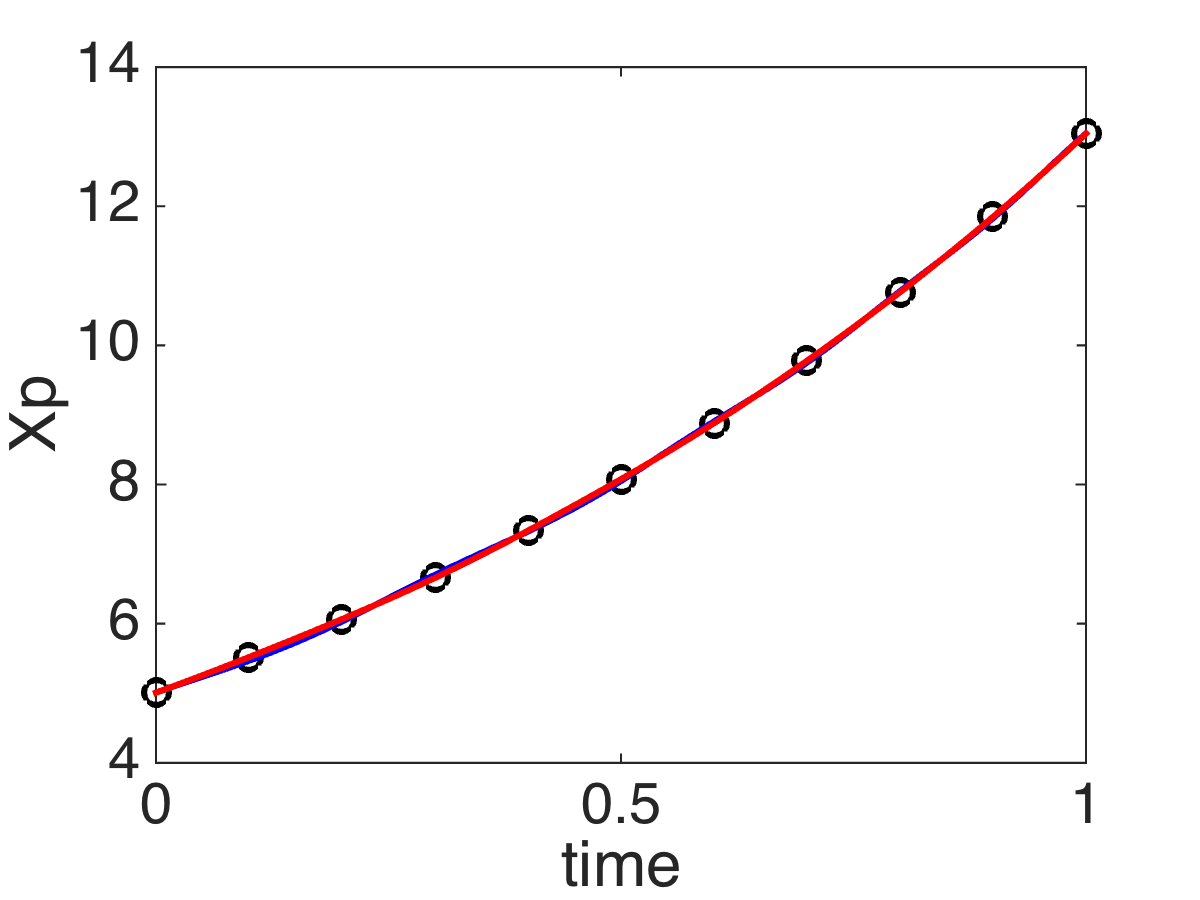}
		\caption{}
		\label{case_3_magnitudes_xp}
	\end{subfigure}
	\caption{Particle (a) velocities and (b) locations using the mean of the exact particle locations as well as the First-Order and SPARSE models with a spatially varying background fluid velocity.}
	\label{case_3_magnitudes} 
\end{figure}

\begin{figure}[ht]
	\centering
	\begin{subfigure}[b]{0.48\textwidth}
		\includegraphics[width=\textwidth]{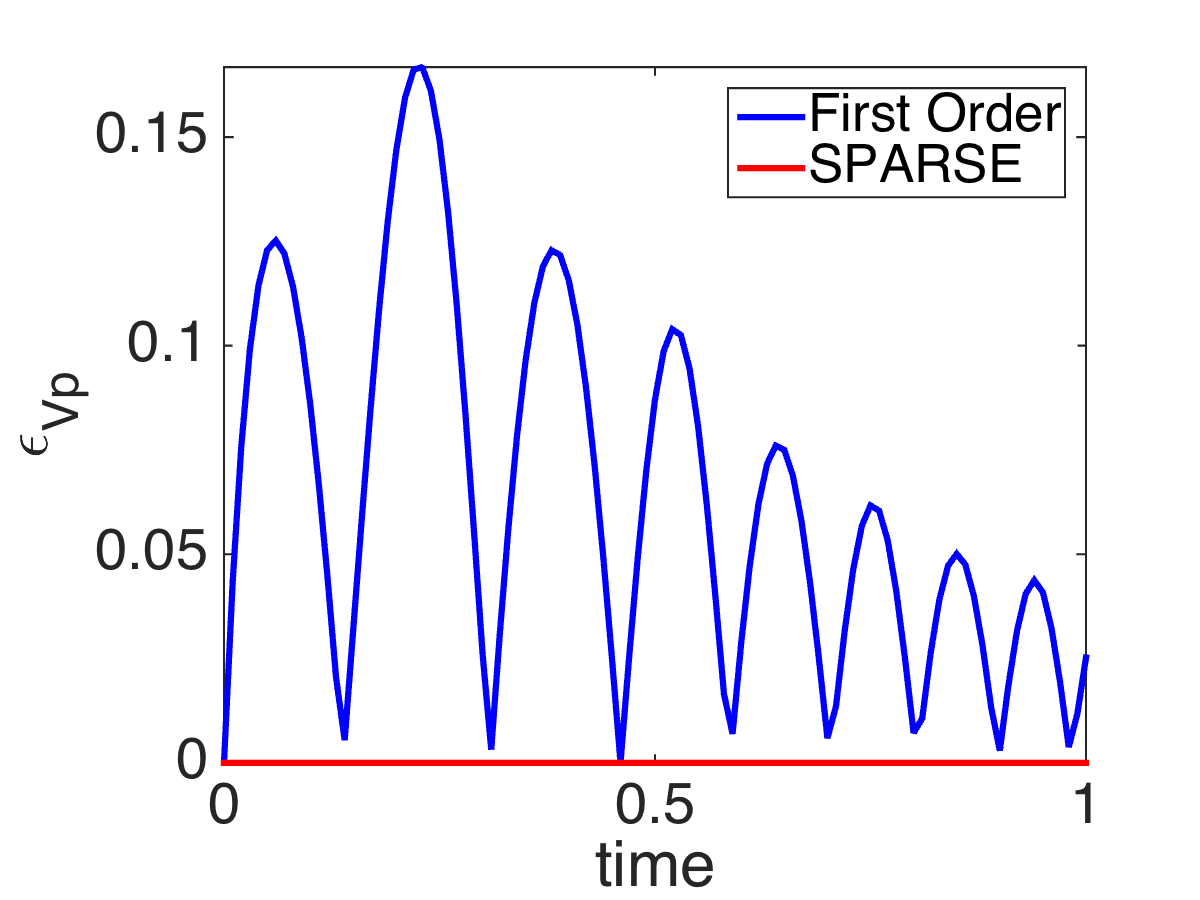}
		\caption{}
		\label{case_3_errors_vp}
	\end{subfigure}
	\begin{subfigure}[b]{0.48\textwidth}
		\includegraphics[width=\textwidth]{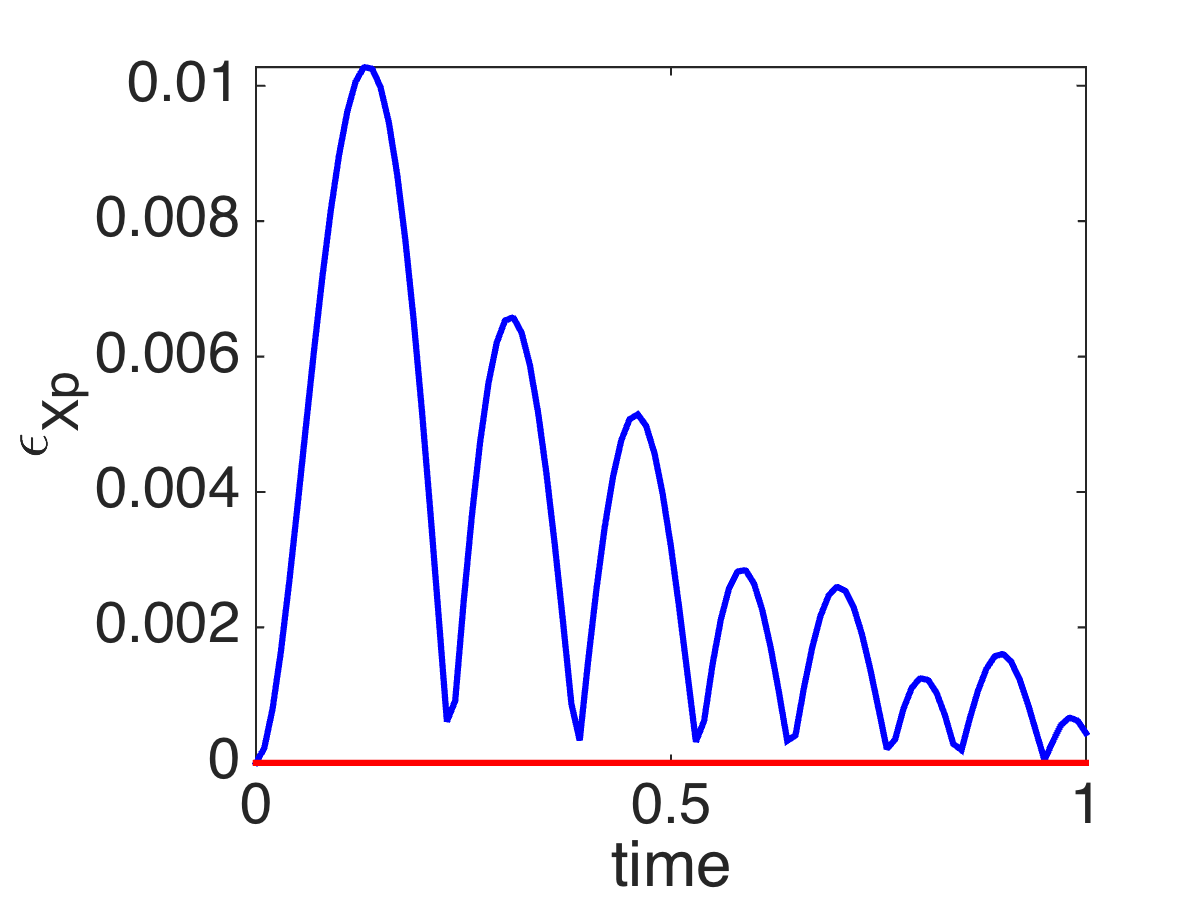}
		\caption{}
		\label{case_3_errors_xp}
	\end{subfigure}
	\caption{Modeling error of the mean particle (a) velocity and (b) location using the First-Order and SPARSE models with a spatially varying background fluid velocity.}
	\label{case_3_errors} 
\end{figure}

Figures \ref{case_3_magnitudes_vp} and \ref{case_3_errors_vp} shows that First-Order modeling of the fluid velocity can
lead to errors as high as 17\% whereas the averaged velocity and location of the cloud is 
\textit{a priori} closed in an exact manner with the SPARSE model.
Because the perturbations in the fluid velocity are periodic, the errors in the modeled particle velocity are oscillatory and only cause a maximum 1\% error in the particle location in Figures
\ref{case_3_magnitudes_xp} and \ref{case_3_errors_xp}.
%

\section{Validation SPARSE:  Decaying Isotropic Turbulence}
\label{3d_isotropic}

To more rigorously test SPARSE a group of particles traced in a 
decaying isotropic turbulence according to the ``exact'' model is compared with the First-Order
model and \textit{a priori} closed SPARSE.

\begin{figure}[h]
	\centering
	\includegraphics[width=\textwidth]{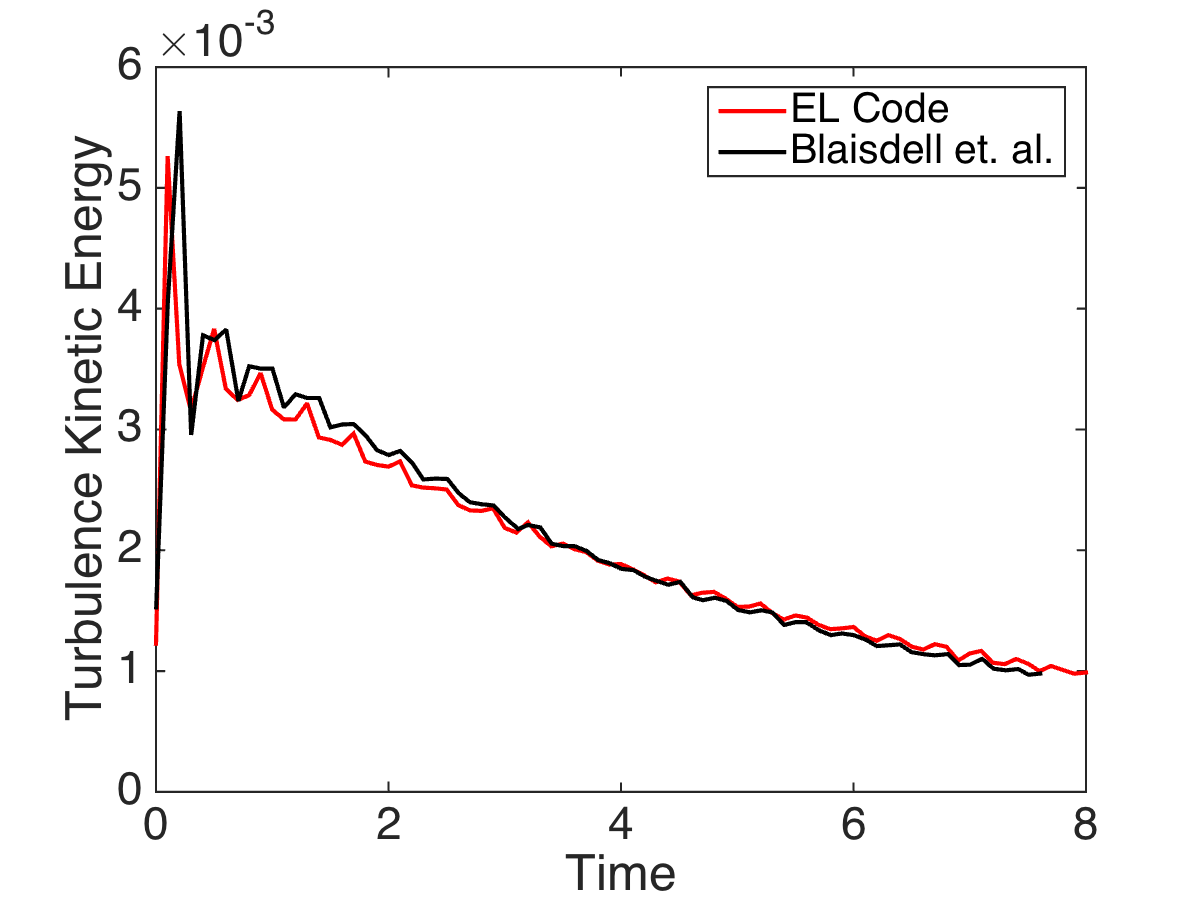}	
	\caption{Comparison of the turbulent kinetic energy (TKE) versus time in isotropic turbulence using a high-order EL code from Jacobs et. al. \cite{JD09} and a Fourier spectral method performed by Blaisdell et. al. \cite{blaisdell_isotropic}.}
	\label{initial_TKE} 
\end{figure}

The isotropic turbulence simulation is performed in a cube with periodic boundary conditions on all sides.
Following \cite{blaisdell_isotropic,JKM05} 
an  initial  correlated flow field is determined based on specified energy spectra.
Computations are performed 
with a compressible Navier-Stokes solver based on a 4$^{th}$ order central difference method for the fluxes 
(as described \cite{JD09}).
To verify the Navier-Stokes solver, we compare computations with this code
with  128 cubed number of grid points 
with the case referred to as "iga96" in \cite{blaisdell_isotropic} computed with 
a $N=96$ Fourier spectral method.
The average initial fluctuating Mach number, 
\begin{align}
M_0 = \frac{\sqrt{\overline{u_i'u_i'}}}{c(\overline{T_0})},
\end{align}
where $u_i'$ is the fluctuating fluid velocity is set to $M_0 = 0.05$.
The reference fluid Reynolds number is set to $Re_f = 2357$.
The decay of turbulent kinetic energy, TKE = $\frac{1}{2}\overline{ {{u_i}^\prime}^2}$, in Figure \ref{initial_TKE}
compares well with the results from Blaisdell et. al. \cite{blaisdell_isotropic}. 
The oscillatory trend in TKE is well-known and documented and is caused by pressure-dilation \cite{kida}.

To test the particle models an exact case that initializes $30^3=27,000$ particles uniformly over 
3x3x3 grid cells,  serves as a reference against which
the First-Order model and SPARSE are tested.
As compared to iga96 the initial velocity field is scaled by  a factor five. This ensures that 
the carrier phase disperses the particle cloud more significantly than the iga96 velocity field 
and hence the effects of the cloud modeling are more visible. 
The initial particle velocity is  set according to a uniform random
number around zero with an absolute maximum of 2.5.
We use the following drag correction function 
\be
C_D = \left(0.38+\frac{24}{Re_p}+\frac{4}{Re_p{\nicefrac{1}{2}}}\right)\left(1+e^{-\frac{0.43}{M_p^{4.67}}}\right),
\label{eq:f1_2}
\ee
that corrects the Stokes drag for high relative particle Reynolds number, $Re_{p}=|v_{f}-v_{p}|d_{p}/\nu$, and 
Mach number, $M_{p}=|v_{f}-v_{p}|/\sqrt{T_{f}}$, according to Boiko et. al. \cite{BKKPPF97, JD09, Davis1, Davis2, Davis_carbuncle}.
The particles' response time is set to $\tau_p = 0.01$.
Matlab was used to compute the partial derivatives of the drag coefficient equation, which are needed for the SPARSE model in
(\ref{eq:SPARSE}).

\begin{figure}[ht]
	\centering
	\begin{subfigure}[b]{0.45\textwidth}
		\includegraphics[width=\textwidth]{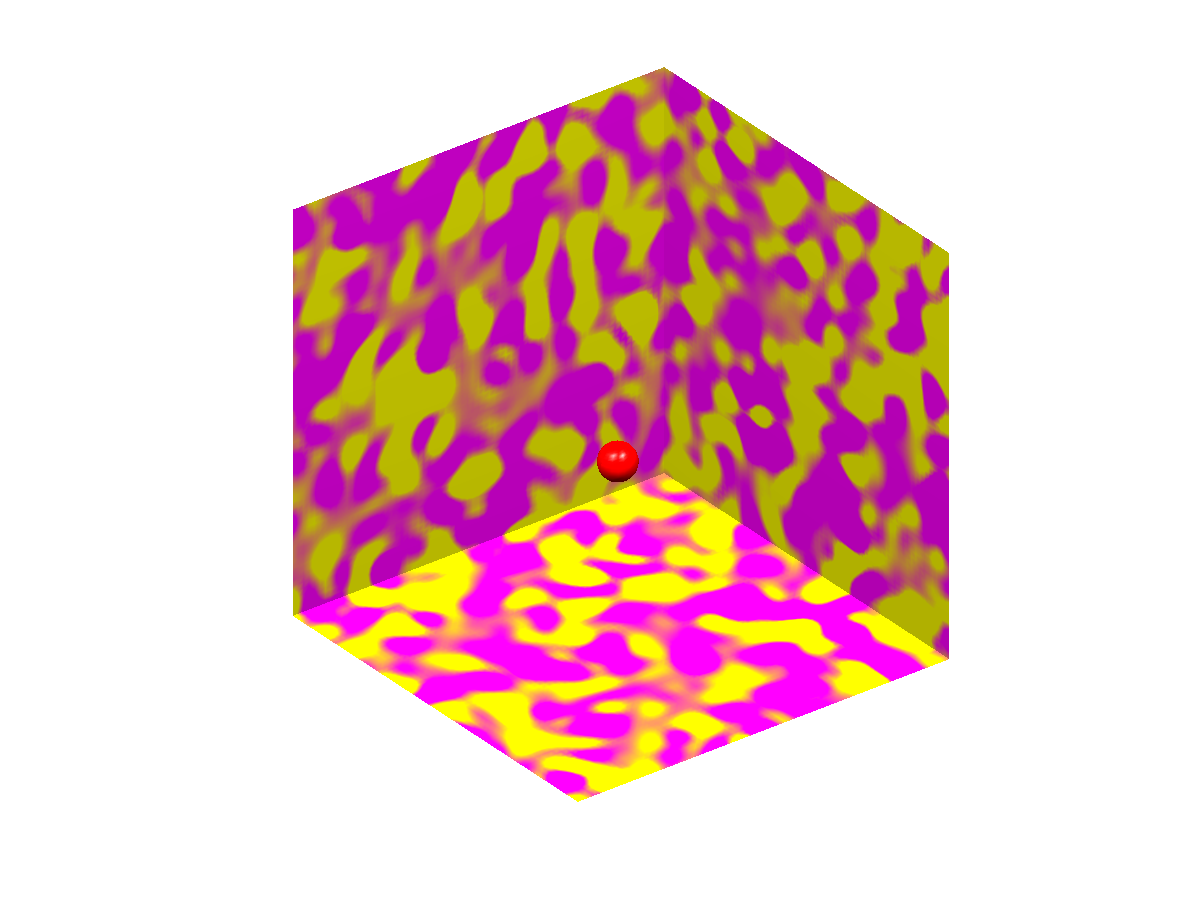}
		\caption{t = 0}
	\end{subfigure}
	\begin{subfigure}[b]{0.45\textwidth}
		\includegraphics[width=\textwidth]{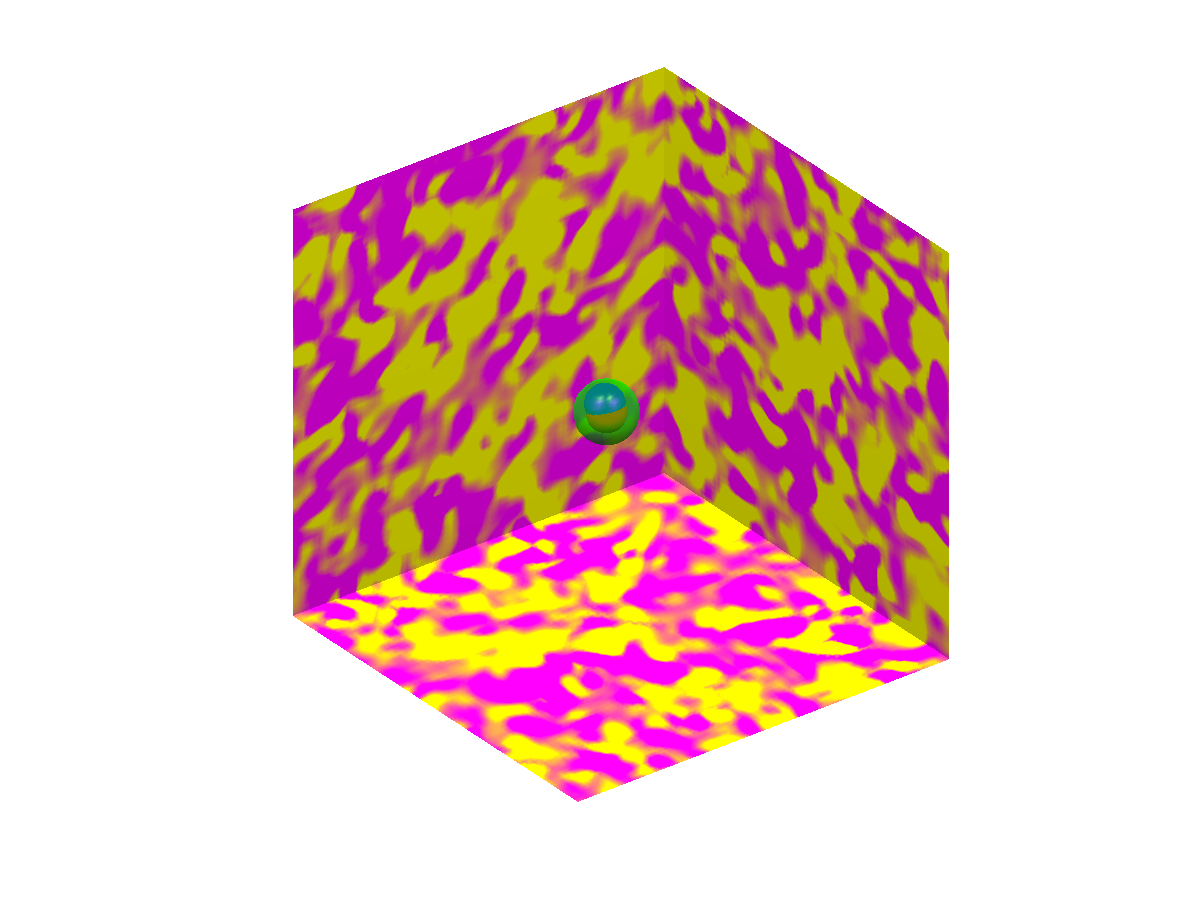}
		\caption{t = 2}
	\end{subfigure}
		\begin{subfigure}[b]{0.45\textwidth}
		\includegraphics[width=\textwidth]{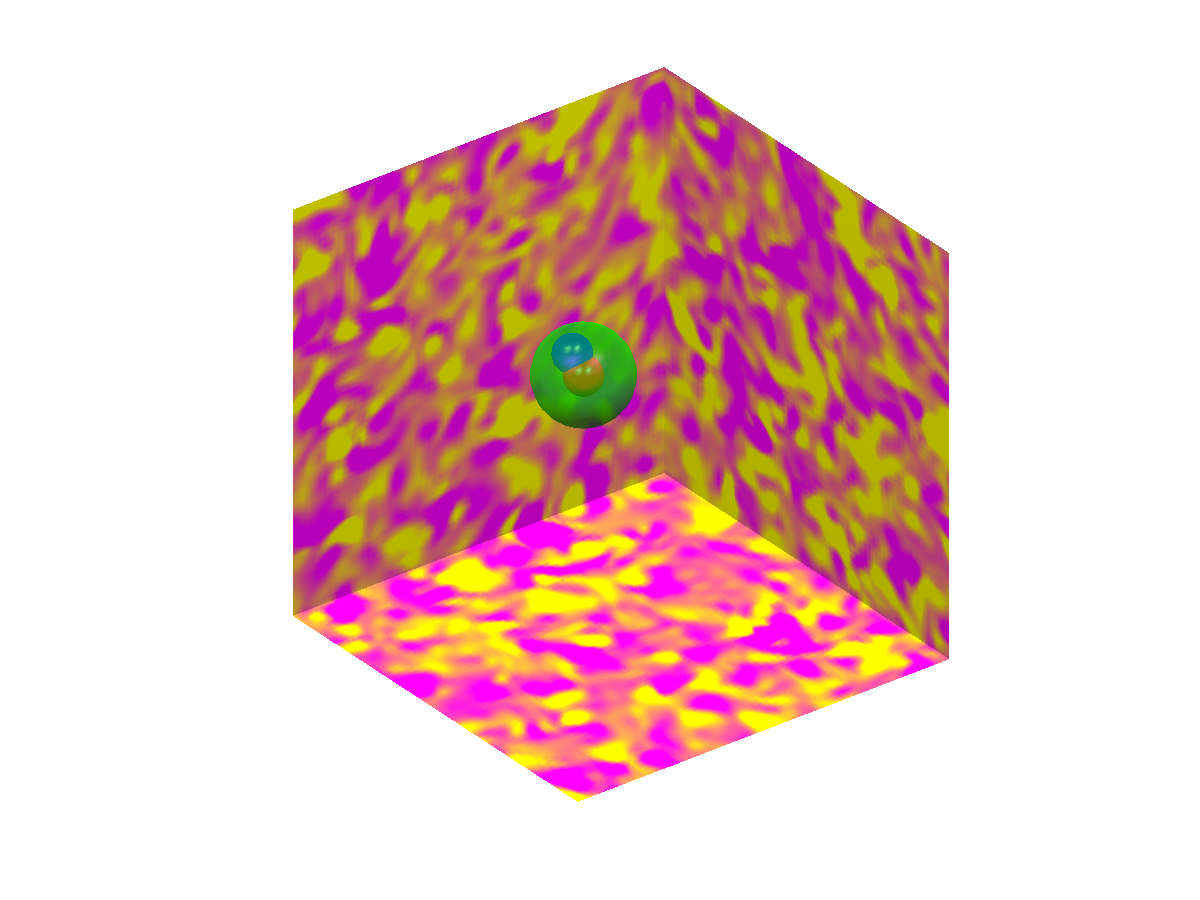}
		\caption{t = 4}
	\end{subfigure}
	\begin{subfigure}[b]{0.45\textwidth}
		\includegraphics[width=\textwidth]{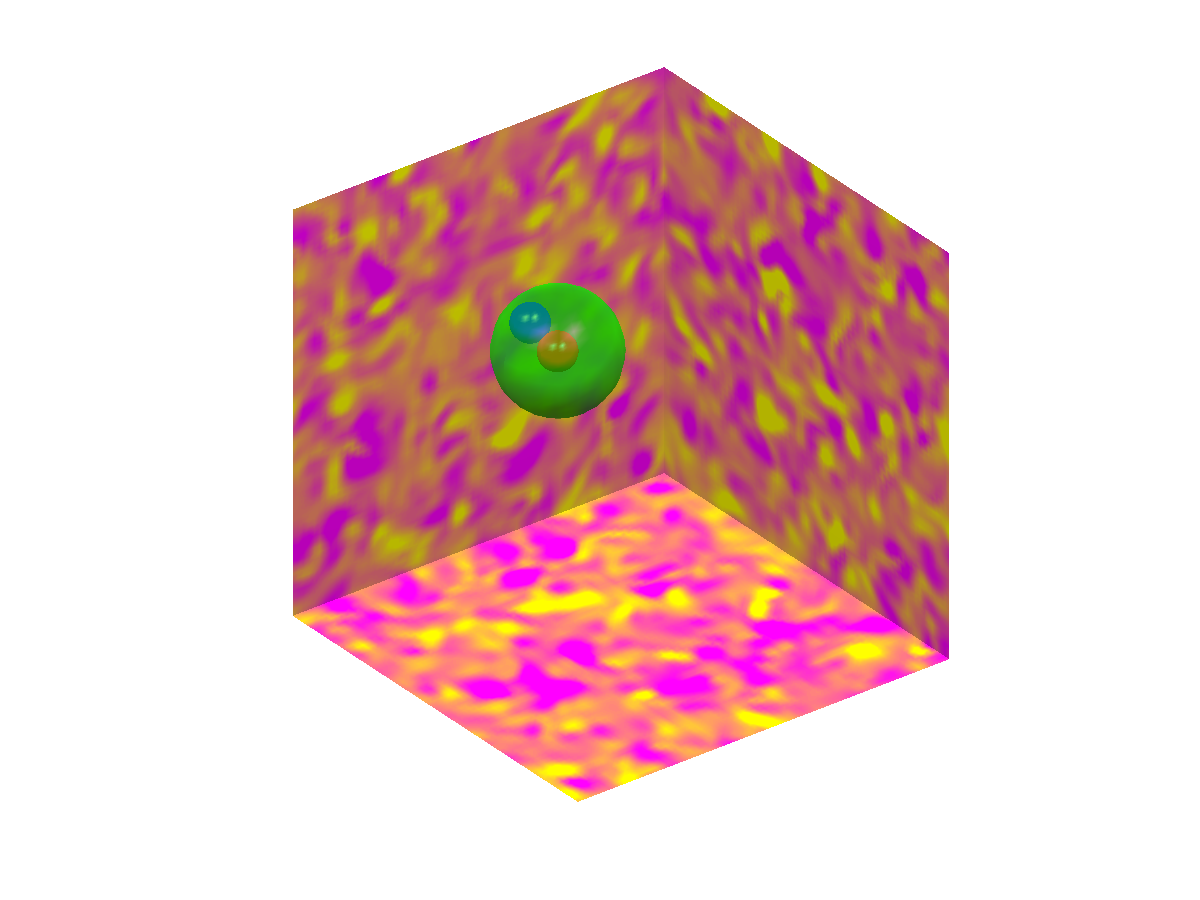}
		\caption{t = 6}
	\end{subfigure}
		\begin{subfigure}[b]{0.45\textwidth}
		\includegraphics[width=\textwidth]{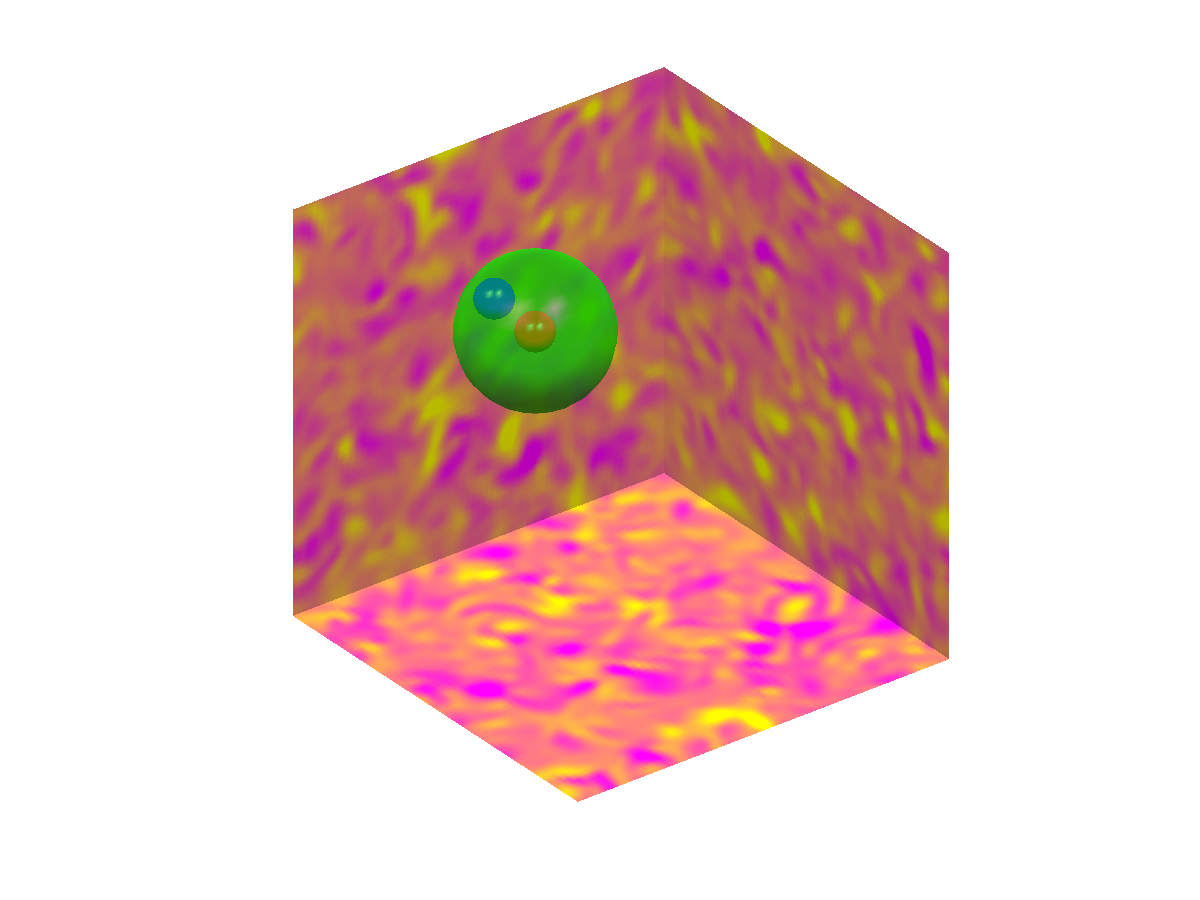}
		\caption{t = 8}
	\end{subfigure}
	\begin{subfigure}[b]{0.45\textwidth}
		\includegraphics[width=\textwidth]{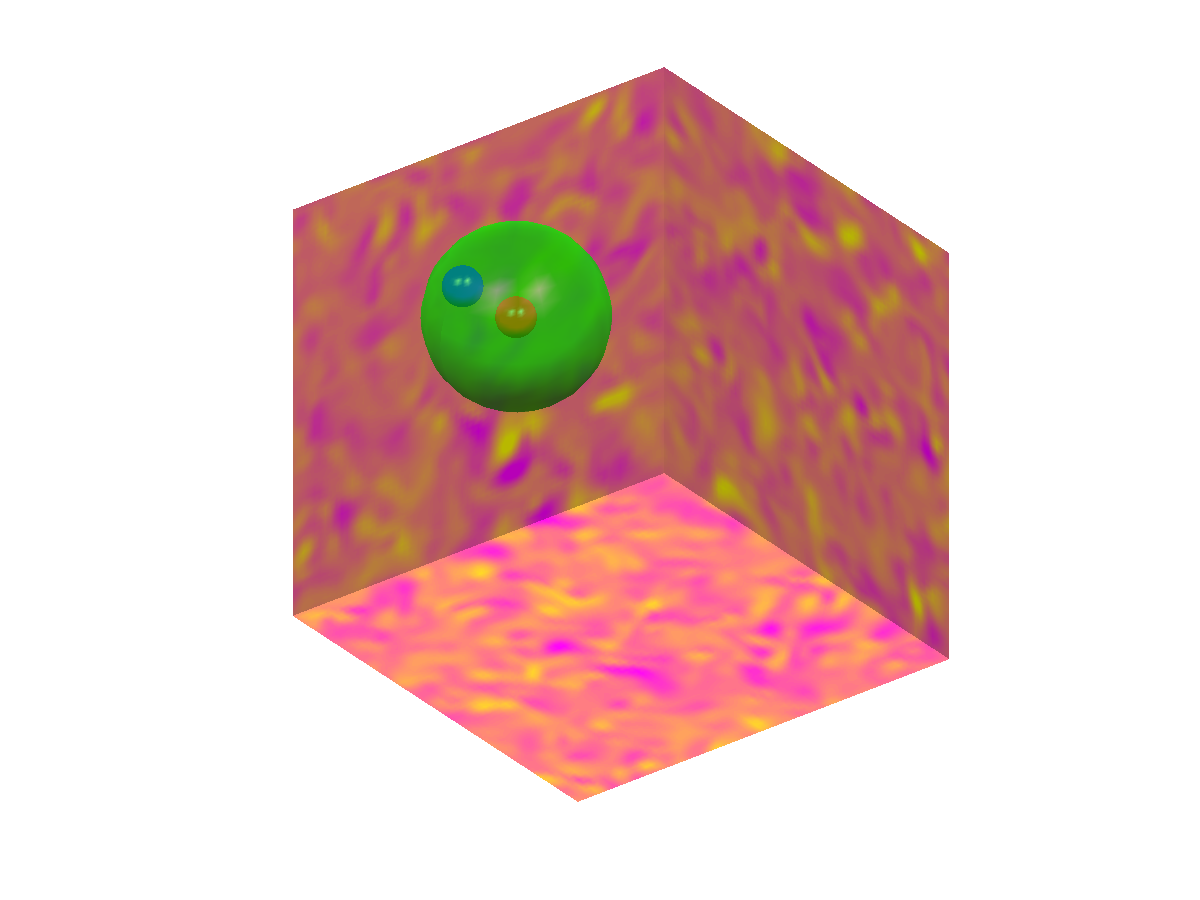}
		\caption{t = 10}
	\end{subfigure}
	\caption{Helicity projected onto the cubical computational domain faces
and particle cloud locations for the 3D isotropic decaying turbulence case at
times t = (a) 0.0, (b) 1.6, (c) 3.2, (d) 4.8, (e) 6.4 and (f) 8.0. The First-Order
 modeled particle is visualized by  a blue sphere, the SPARSE modeled by a
red (sphere) and the exact model by a  green sphere. The root mean square
dispersion of the the exact location is visualized by the diameter of the
green sphere. The  diameter of the red and blue sphere visualize the initial
rms dispersion.} \label{3D_Particles} 
\end{figure}

The time lapse in Figure \ref{3D_Particles} 
shows that the  cloud modeled with SPARSE (red sphere)
closely follows the average location of the exact cloud (green sphere), while
the First-Order model (red sphere) deviates significantly from the exact cloud.
The temporal dispersion of the exact cloud is visualized by the large  diameter of the 
exact model (green) sphere as compared to initial diameter of the (red and blue) cloud approaches.
Dispersion, as measured by for example the rate of change of the cloud radius, is not modeled in the  SPARSE  and First-Order model and
it is the subject of ongoing investigation.

\begin{figure}[ht]
	\centering
	\begin{subfigure}[b]{0.45\textwidth}
		\includegraphics[width=\textwidth]{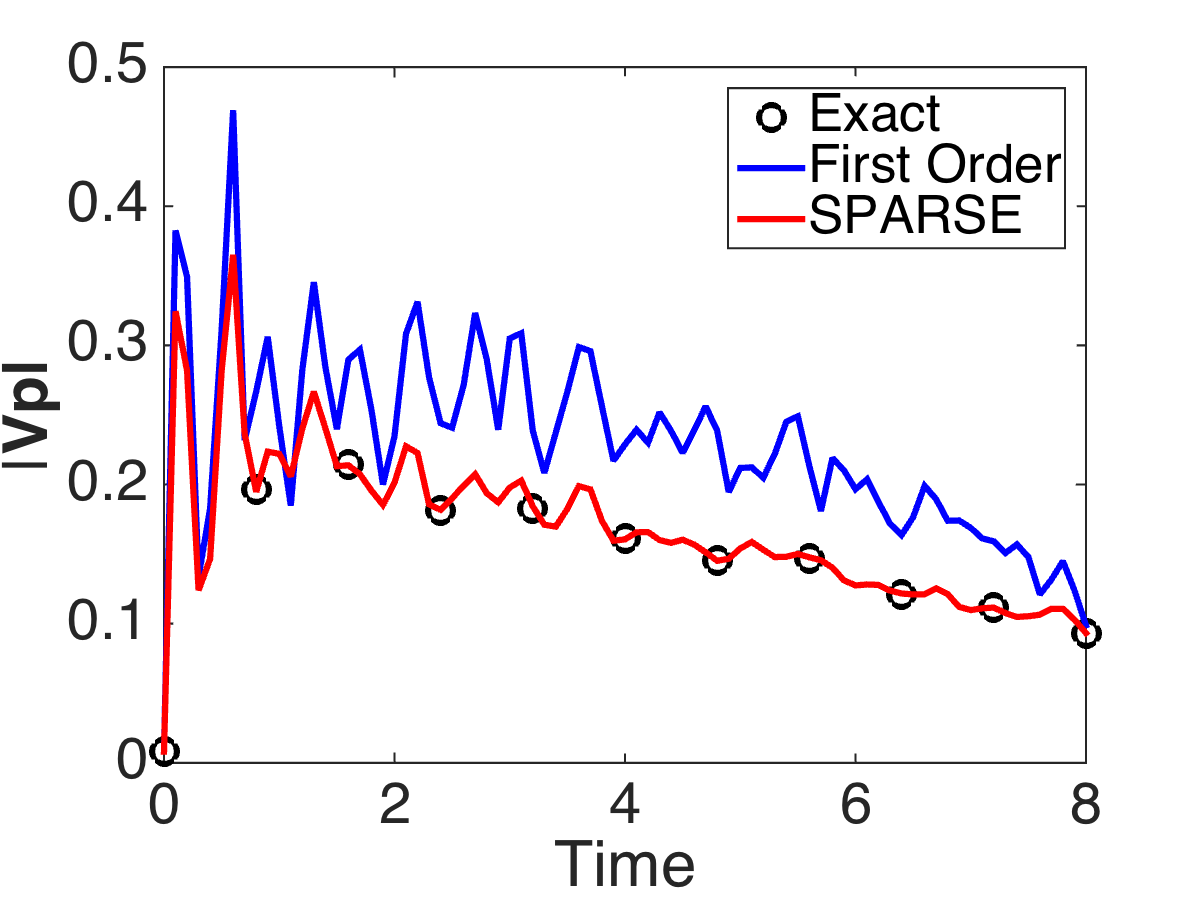}
		\caption{}
	\end{subfigure}
	\begin{subfigure}[b]{0.45\textwidth}
		\includegraphics[width=\textwidth]{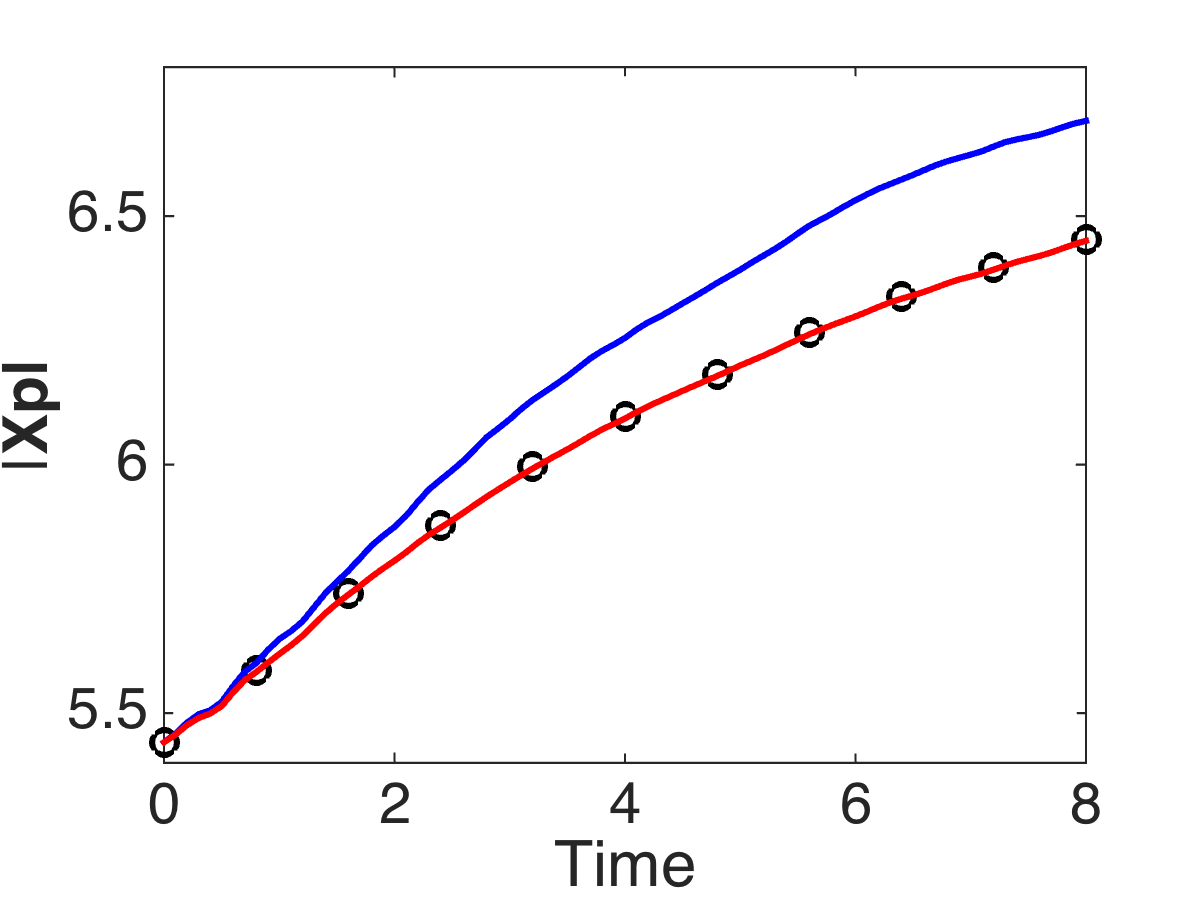}
		\caption{}
	\end{subfigure}
	\caption{The magnitude of the average (a) velocity and (b) particle distance from the origin is shown when computed using the 
First-Order model, SPARSE model and average over the physical particles.}
	\label{avg_stats_3d} 
\end{figure}

A comparison of the average particle cloud distance from the origin 
\begin{align}
|{\bf x_p}| = \sqrt{x_{p}^2+y_{p}^2+z_{p}^2},\nonumber
\end{align}
and  averaged velocity, 
\begin{align}
|{\bf v_p}| = \sqrt{u_{p}^2+v_{p}^2+w_{p}^2},
\end{align}
in Figure \ref{avg_stats_3d} confirms the improved modeling by SPARSE as compared to the First-Order model.
While both the exact cloud and the computational clouds are initially entrained
 in the same turbulent eddy, the average velocity over the cloud used in SPARSE is near zero, while the local velocity used
in the First-Order model is large. The latter cloud is hence displaced more.
With increasing time, the exact cloud disperses  and the fluid velocity is sampled over a larger area. 
Because the turbulence is isotropic and decaying the averaged  fluid velocity goes to zero over a larger
area and increasing time, respectively. The local velocity decays in time also, but deviates from the near zero, averaged
velocity leading to the deviation of the First-Order model in time.

\begin{figure}[ht]
	\centering
	\begin{subfigure}[b]{0.45\textwidth}
		\includegraphics[width=\textwidth]{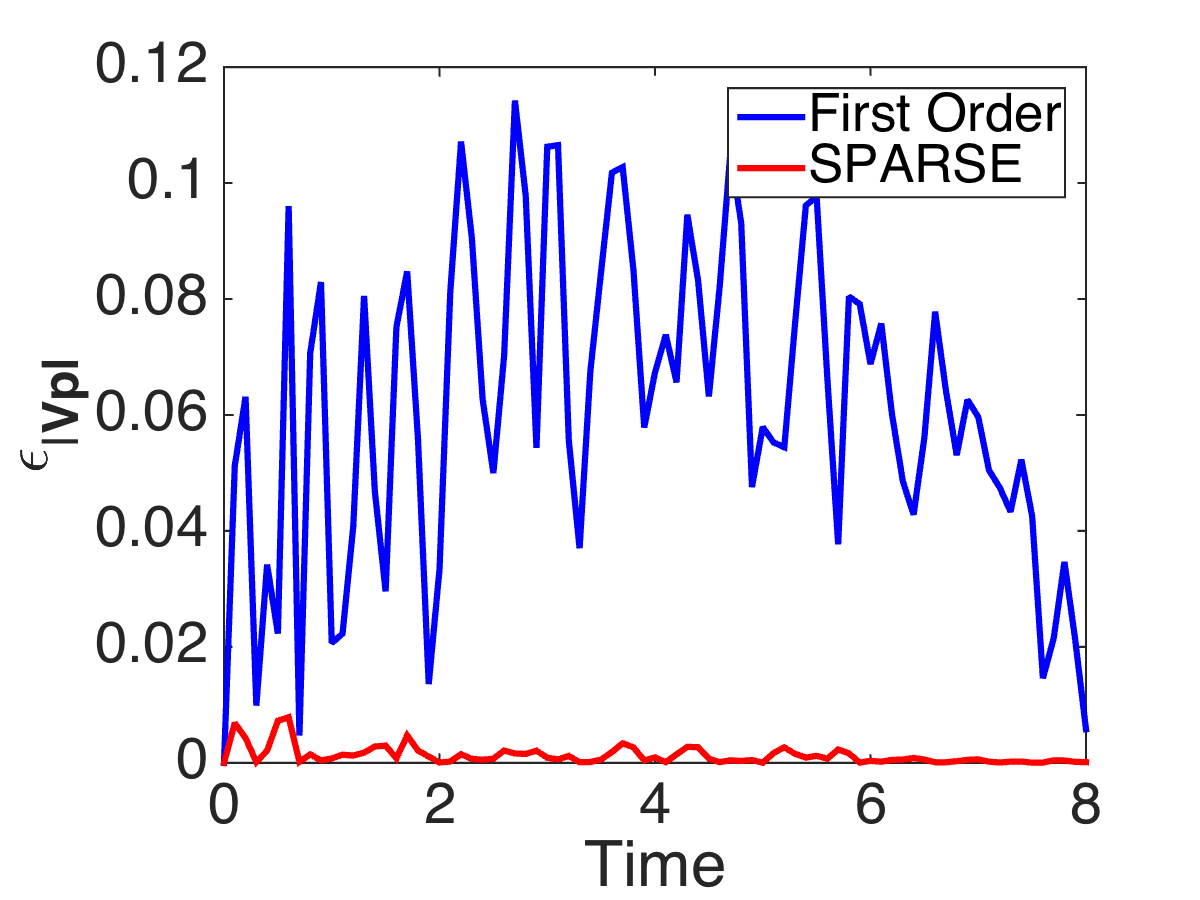}
		\caption{}
	\end{subfigure}
	\begin{subfigure}[b]{0.45\textwidth}
		\includegraphics[width=\textwidth]{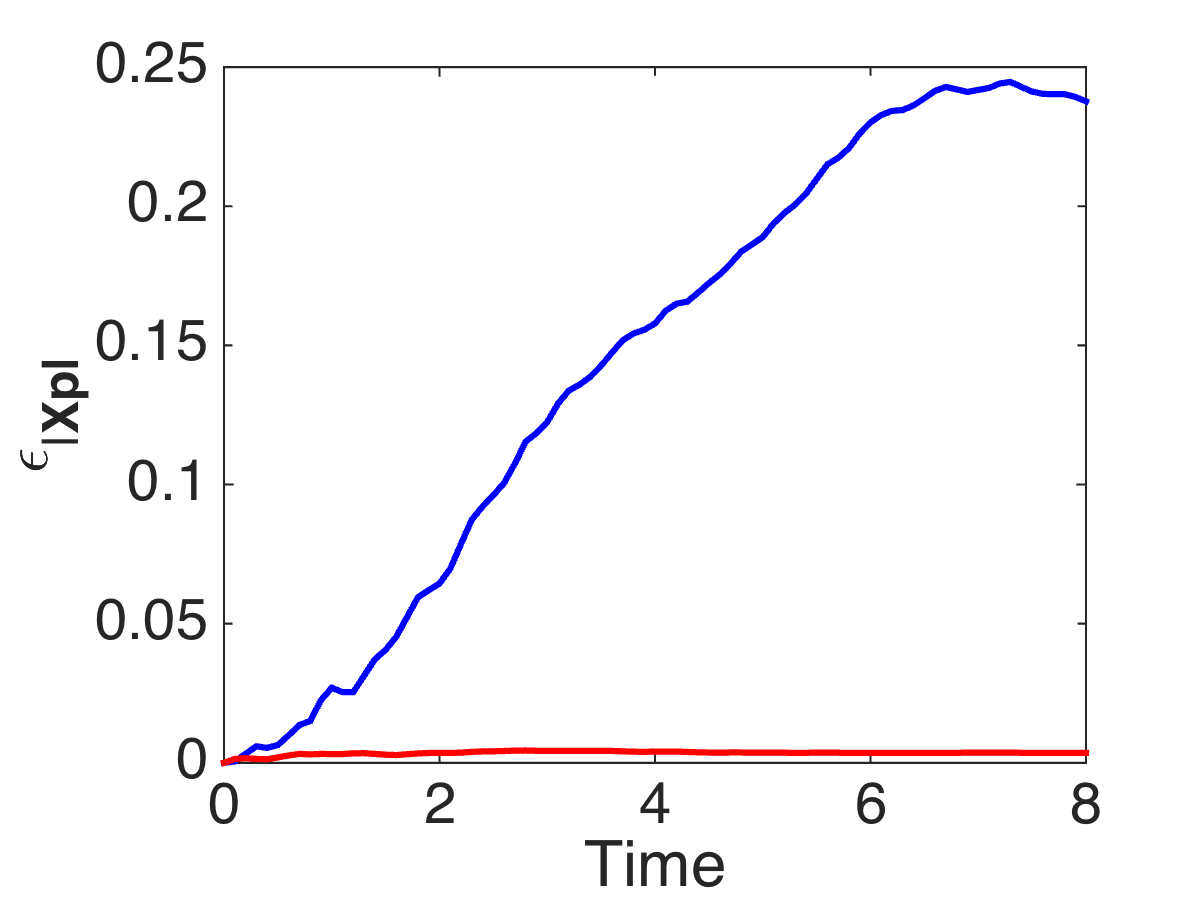}
		\caption{}
	\end{subfigure}
ma	\caption{The error in computing the magnitude of the average (a) velocity and (b) particle distance from the origin is shown when computed using the First-Order model, SPARSE model and average over the physical particles.}
	\label{avg_errors_3d} 
\end{figure}

Errors in the magnitude of the distance from the origin and velocity plotted
in Figure \ref{avg_errors_3d} shows that SPARSE modeling errors are non-zero, because the drag correction in  
(\ref{eq:f1_2})is
non-linear and hence the Taylor truncation error in (\ref{taylor_expansion}) used for SPARSE is non-zero. 
The error however is small, within 0.5\% of the exact model. 

\begin{figure}[ht]
	\centering
		\includegraphics[width=0.45\textwidth]{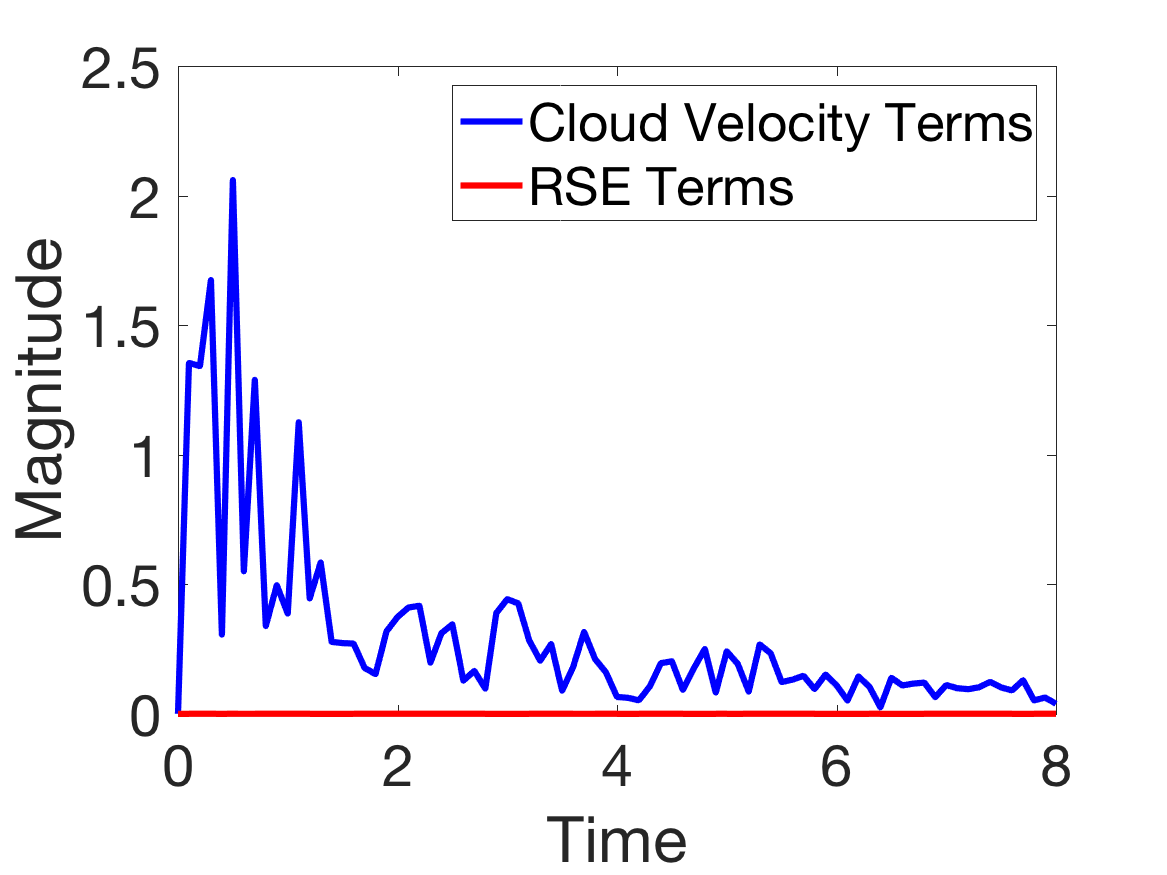}
		\caption{}
		\label{}
	\caption{Comparison of the two terms on the right hand side of (\ref{eq:SPARSE},
$f(\bar{\bf a})\bar{\bf a}$ and $\frac{df(\bar{\bf a})}{d {\bf a}}\overline{{\bf a}^\prime {\bf a}^\prime}$. The first term (blue line) is dependend on the SPARSE fluid velocity modeling and the second term (red line) is proportional to the interphase stresses. 
}
	\label{RHS_comp} 
\end{figure}

The impact of  two modeling components of the SPARSE model, including the averaging of the fluid velocity of the cloud  and the stress modeling are compared
in Figure \ref{RHS_comp}. Plotted are the  two terms on the right-hand side of
(\ref{eq:SPARSE}). The first term that is affected by the cloud velocity averaging 
is significantly larger than the second term that is proportional to the sub-cloud stresses. The second term is on the order of $10^{-5}$. In general, the effect of the  sub-cloud stresses is not necesarilly smaller than the effect of velocity averaging as was seen in the 1D tests.

\clearpage

\section{Conclusions and Future Work}
\label{conclusion}

SPARSE provides a model that traces a group of point particles through a single computational point parcel.
The model improves upon Cloud-In-Cell methods by, firstly, accounting for fluid and particle  stress
(i.e. the Reynolds stress equivalent) terms 
and, secondly, averaging of the fluid velocity at the Lagrangian computational particle.
One-dimensional and three-dimensional \textit{a priori} tests show that both improvements yield
an excellent comparison of the  averaged trace of a cloud of particles and the computational parcel.

This paper is only  a first step in the development of a closed SPARSE model. Current investigation
focuses on closure based on a multi-scale approach as reported in \cite{Sen15}.

\section*{Acknowledgements}
We gratefully acknowledge the financial support by the Air Force Office of Scientific Research under grant number FA9550-16-1-0008.

\bibliographystyle{elsarticle-num}
\bibliography{JMF_Bib}

\end{document}